\documentclass[12pt]{article}

\usepackage{cancel}
\usepackage{amssymb,amsmath}
\usepackage[usenames,dvipsnames]{color}
\usepackage[implicit=true, 
  colorlinks=true,linkcolor=Blue,citecolor=PineGreen,urlcolor=BrickRed]{hyperref}

\def\harr#1#2{\smash{\mathop{\hbox to .3in{\rightarrowfill}}
 \limits^{\scriptstyle#1}_{\scriptstyle#2}}}

\def\s2{\frac{1}{\sqrt2}}

\def\be{\begin{equation}}
\def\ee{\end{equation}}
\def\beqa{\begin{eqnarray}}
\def\eeqa{\end{eqnarray}}

\def\Dsl{\,\raise.15ex\hbox{/}\mkern-13.5mu D} 
\def\d3{d^3}

\def\IR{\mathbb{R}}

\def\IZ{\mathbb{Z}}

\def\R{\mathbb{R}}
\def\C{\mathbb{C}}
\def\Z{\mathbb{Z}}

\def\Q{\mathbb{Q}}

\def\H{\mathbb{H}}

\begin{document}

\vspace{.5cm}
\begin{center}
\Large{\bf Invariants of Four-Manifolds with Flows Via Cohomological
Field Theory}\\
{\normalsize Dedicated to John Milnor on his $80^{\rm th}$ birthday}\\
\vspace{1cm}

\large Hugo Garc\'{\i}a-Compe\'an$^a$\footnote{e-mail address: {\tt
compean@fis.cinvestav.mx}}, Roberto Santos-Silva$^a$\footnote{e-mail address: {\tt
rsantos@fis.cinvestav.mx}}, Alberto Verjovsky$^b$\footnote{e-mail address: {\tt
alberto@matcuer.unam.mx}}
\\
[2mm] {\small \em $^a$Departamento de F\'{\i}sica, Centro de
Investigaci\'on y de
Estudios Avanzados del IPN}\\
{\small\em P.O. Box 14-740, 07000 M\'exico D.F., M\'exico}
\\[4mm]
{\small \em $^b$Instituto de Matem\'aticas, UNAM, Unidad Cuernavaca}\\
{\small\em Av. Universidad s/n, Col. Lomas de Chamilpa}\\
{\small\em c.p. 62210, Cuernavaca Morelos, M\'exico}
\\[4mm]

\vspace*{1.5cm}
\small{\bf Abstract} \\
\end{center}

\begin{center}
\begin{minipage}[h]{14.0cm} { The Jones-Witten invariants can be generalized
for non-singular smooth vector fields with invariant probability
measure on $3$-manifolds, giving rise to new invariants of dynamical
systems \cite{VV}. After a short survey of cohomological field
theory for Yang-Mills fields, Donaldson-Witten invariants are
generalized to four-dimensional manifolds with non-singular smooth
flows generated by homologically non-trivial $p$-vector fields.
These invariants have the information of the flows and they are
interpreted as the intersection number of these flow orbits and
constitute invariants of smooth four-manifolds admitting global
flows. We study the case of K\"ahler manifolds by using the Witten's
consideration of the strong coupling dynamics of ${\cal N}=1$
supersymmetric Yang-Mills theories. The whole construction is
performed by implementing the notion of higher dimensional
asymptotic cycles \`a la Schwartzman \cite{SH}. In the process
Seiberg-Witten invariants are also described within this context.
Finally, we give an interpretation of our asymptotic observables of
4-manifolds in the context of string theory with flows.}
\end{minipage}
\end{center}

\bigskip

\date{\today}

\vspace{3cm}


\newpage

\section{Introduction}

Quantum field theory is not only a framework to describe the physics
of elementary particles and condensed matter systems but it has been
useful to describe mathematical structures and their subtle
interrelations. One of the most famous examples is perhaps the
description of knot and link invariants through the correlation
functions of products of Wilson line operators in the Chern-Simons
gauge theory \cite{WJP}. These invariants are the Jones-Witten
invariants or Vassiliev invariants depending on whether the coupling constant
is weak or strong respectively. Very recently some aspects of
gauge and string theories found a strong relation with Khovanov
homology \cite{Witten:2011zz}.

In four dimensions the Donaldson invariants are invariants of the
smooth structure on a closed four-manifold. This is in the sense
that if two homeomorphic differentiable manifolds have different
Donaldson invariants then they are not diffeomorphic
\cite{Donaldson,DK}. These invariants were reinterpreted by Witten
in terms of the correlation functions of suitable observables of a
cohomological Yang-Mills field theory in four dimensions \cite{WTQ}.
Such a theory can be obtained from an appropriate topological twist
on the global symmetries of the ${\cal N}=2$  supersymmetric
Yang-Mills theory in Minkowski space with global $R$-symmetry
SU$(2)$ that rotates the supercharges. A gravitational analog of the
Donaldson theory is given by the topological gravity in four and two
dimensions \cite{WTG}. The computation of Donaldson invariants for
K\"ahler manifolds has been done from the mathematical point of view
in Refs. \cite{GRADY,KM}. These Donaldson invariants were later
reproduced in Ref. \cite{MINIMAL} by using the strong coupling
dynamics of ${\cal N}=1$ supersymmetric gauge theories in four
dimensions. Precisely a deeper understanding of the dynamics of
strong coupling ${\cal N}=2$ supersymmetric Yang-Mills theories in
four dimensions \cite{SW} including the notion of $S$-duality,
allowed to give an alternative approach to Donaldson theory in terms
of the low energy effective abelian gauge theory coupled to magnetic
monopoles \cite{MONO}. For a recent account of all these
developments, see \cite{book1}.

Moreover, the topological twist was applied to other theories such as
string theory resulting in the so called topological sigma models
\cite{TSM}. The two possible twists of the global symmetries of the
world-sheet theory leads to the so called $A$ and $B$-models, whose
correlation functions give rise to a description of the moduli space
in terms of only the K\"ahler cone or only the moduli of complex
structures of a target space Calabi-Yau manifold. $A$-models give
rise to Gromov-Witten invariants. Mirror symmetry is realized
through the interchanging of $A$ and $B$ models of two Calabi-Yau
manifolds related by the interchanging of Betti numbers \cite{MM}.
For a recent survey of all these topological field theories and
their interrelations, see for instance \cite{Marino:2005sj}.

On the other hand it is well known that topology and symplectic
geometry play a very important role in the theory of dynamical
systems \cite{khesin}. Schwartzman introduced some years ago,
homology 1-cycles associated to a foliation known as {\it asymptotic
cycles} \cite{SA}. These 1-cycles are genuine homology cycles and
they represent an important tool to study some properties of
dynamical systems. Moreover, the generalization to $p$-cycles, with
$p>1$, was done in Ref. \cite{SH}. Such generalization was achieved
by using some concepts of dynamical systems such as flow boxes and
geometric currents \cite{RS,S}. The definition of asymptotic cycles
for non-compact spaces was discussed in Ref. \cite{NCS}. In
particular, the results \cite{SA} were used define the
Jones-Witten polynomial for a dynamical system \cite{VV}. More
recently the ideas from \cite{SH,RS,S} were used to find new
suitable higher dimensional generalizations of the asymptotic
linking number starting from a topological $BF$ theory (see
\cite{GarciaCompean:2009zh} and references therein).

In the present paper we also use the notion of asymptotic $p$-cycles
to extend the Donaldson-Witten and Seiberg-Witten invariants when
smooth $p$-vector fields are incorporated globally on the underlying
four-manifold. The asymptotic $p$-cycles associated to $p$-vectors
on the manifolds define real homology $p$-cycles on these manifolds.
They will constitute refined topological invariants of dynamical
systems which distinguish the triplet $(M,{\cal F},\mu)$, where $M$
is a four-manifold, ${\cal F}$ is the foliation (possibly singular)
associated to a $p$-vector and $\mu$ is a transverse measure of
${\cal F}$ which is invariant under holonomy.
Two triplets $(M_1,{\cal F}_1,\mu_1)$ and $(M_2,{\cal
F}_2,\mu_2)$ are {\it differentiably equivalent} if there is a
diffeomorphism from $M_1$ to $M_2$, which sends the leaves of ${\cal
F}_1$ to the leaves of ${\cal M}_2$ and the push-forward of $\mu_1$
is $\mu_2$. Moreover these invariants will constitute a
generalization of the Donaldson-Witten invariants for such triplets. For
instance, one of the main results here is that our invariants will
distinguish triplets: if two triplets  $(M_i,{\cal F}_i,\mu_i)$
($i=1,2$) have the property that the four-dimensional
Donaldson-Witten invariants of $M_1$ and $M_2$ are equal but our
invariants are different then the corresponding systems of flows on
them are not differentiably equivalent.

On the other hand it is well known that Donaldson-Witten invariants
can be interpreted in terms of the scattering amplitude (at zero
momentum) of an axion with a NS5-brane in the heterotic string
theory \cite{Harvey:1991hq}. This paper would suggest a possible
physical interpretation of our invariants involving flows in terms
of an averaged propagation of a closed string in a target space
described in terms of the moduli space of positions of  a NS5-brane.
That means, a ``continuous'' flux of closed strings (propagating in
the transverse space to the worldvolume of the NS5 brane) giving
rise to an asymptotic 2-cycle. The diffuseness of the asymptotic
cycle is determined by a flow (or set of flows) in the target space
given by some field in the target space, for instance, the NS
$B$-field whose associated 2-vector field gives the 2-foliation on
the target.

The organization of the present paper is as follows: Sec. 2 is
devoted to a brief review of asymptotic $p$-cycles with $p>1$. In
Sec. 3 we overview cohomological field theory for Donaldson-Witten
theory. In Sec. 4 we define the Donaldson-Witten invariant for
four-dimensional manifolds in the presence of a smooth and nowhere
vanishing $p$-vector field over the underlying spacetime manifold.
It is also verified that this invariant is well defined as a
limiting average of the standard definition. Section 5 is devoted to
describe the procedure for K\"ahler four-manifolds. This is done by
using a physical procedure through the incorporation of a mass term
which breaks the supersymmetry to ${\cal N}=1$ theories allowing the
existence of a mass gap. In Section 6 we survey the Seiberg-Witten
invariants. We focus mainly on the case of abelian magnetic
monopoles. Non-abelian monopoles are also briefly described. In Sec.
7 we derive the Seiberg-Witten invariants in the presence of flows.
Sec. 8 is devoted to explain how the Donaldson-Witten invariants for
flows can be derived from a suitable system of strings in
non-trivial flows on the spacetime target space. Finally, in Sec. 9
our final remarks and conclusions close the paper.

\section{Asymptotic Cycles and Currents}

In this section we give a brief overview of asymptotic $p$-cycles
with $p \geq 1$. Our aim is not to provide an extensive review of
this material but introduce the notations and conventions of the
relevant structures, which will be needed in the subsequent
sections. For a more complete treatment see Refs.
\cite{SA,SH,RS,S,dRham}.

In order to study the main aim of the paper, which is a
generalization of invariants of four-manifolds in the presence of a
non-singular flows over a closed four-di\-men\-sional manifold $M$,
it is necessary to consider asymptotic homology $p$-cycles of the
flow on $M$ with values of $p$ greater than one. Here we will have
two possibilities. The first one corresponds to a flow generated by
a $p$-vector field which is not localized in the homology $p$-cycles
of $M$. The second possibility is when the $p$-vector field is
defined only on the tangent space of the $p$-cycles of $M$. Of
course we could have a mixed situation. We also consider a set of
flow invariant probability measures supported on the whole
underlying manifold $M$. In this case the cycles constitute some
``diffuse'' cycles depending on the flow and the measure. The
invariants constructed from these cycles detecting the
differentiable structure of the four-manifolds with flows will be
the asymptotic polynomial invariants of $M$. These invariants will
coincide with the standard Donaldson-Witten invariants when the
measure set is supported on the homology $p$-cycles $\gamma_p$ of
$M$. For simply connected closed 4-manifolds we will be interested
in cycles of dimension $p=0,2,4$. From physical reasons $p=4$ is not
an interesting case since it gives a topological term that can be
added to the classical Lagrangian while that for $p=0$ it is a
trivial cycle. Thus the only relevant cycle will be for $p=2$.  In
this section we define and interpret the observables as currents in
terms of the winding number of asymptotic cycles.

The case $p=1$ was discussed in detail by Schwartzman in Ref.
\cite{SA}. In Ref. \cite{VV} asymptotic cycles were applied to the
Jones-Witten theory in order to find refined invariants of dynamical
systems. Recently, these ideas were generalized to higher dimensions with foliations of dimension
grater than one
using the $BF$ theory without a cosmological constant in
\cite{GarciaCompean:2009zh}.

A current on a compact manifold $M$ of dimension $n$, is a linear and
continuous functional in the de Rham complex $\Omega^*(M)$ i.e., satisfying:
\begin{equation}
\label{current}
C [a_1 \omega_1 + a_2 \omega_2 ] = a_1 C[ \omega_1] + a_2 C [\omega_2],
\end{equation}
for all $\omega_1$ and $\omega_2$ differential forms and $a_1$ and
$a_2$ scalars. As an example we define the following current
$\gamma_p [\omega] = \int_{\gamma_p} \omega$, where $\gamma_p$ is a
$p$-cycle of $M$ and $\omega$ is a $p$-form on $M$. Moreover a
closed $(n-p)$-form $\alpha$ also defines a current in the following
way
\begin{equation}
\alpha[\omega] = \int_M \alpha \wedge \omega.
\end{equation}
Another example is the contraction of a $p$-vector field $\nu_p$ and
a $p$-form $\omega$. Let $\nu_p = \nu^{i_1 \ldots i_p}
\partial_{i_1}\wedge \cdots \wedge \partial_{i_p}$ be a $p$-vector
field and $\omega = \omega_{i_1 \ldots i_p} dx^{i_1} \wedge \cdots
\wedge dx^{i_p}$, then we have

\begin{equation}
\nu_p [\omega] = \omega_{i_1 \ldots i_p} \nu^{i_1 \ldots i_p}.
\end{equation}
A current restricted to the space of smooth $m$-forms is called an
$m$-current. Let ${\mathcal D}_m$ denote the topological vector
space (with the weak* topology ) of $m$-currents. Then in Ref.
\cite{dRham} there were constructed a series of boundary operators
$\partial_m:{\mathcal D}_m\to {\mathcal D}_{m-1}$, defined on
arbitrary $m$-currents, which define a chain complex and thus a
homology theory which is dual to the de Rham cohomology.

\subsection{$p$-Cycles and Geometrical Currents}

The definition of asymptotic cycles for higher dimensional
foliations (under suitable hypothesis) starts by considering a
closed subset ${\cal S}$ of a $n$-dimensional manifold $M$, a family
of submanifolds $L_\alpha$ of dimension $p$, such that ${\cal S}=
\cup_\alpha L_\alpha$ defines a {\it partial foliation} ${\cal F}_p$
(or {\it lamination} \cite{Candel}, see chapter 10) of dimension
$p$. If $M$ is compact we cover all $M$ (including the interior)
with a finite collection of closed disks ${\bf D}^p \times {\bf
D}^{n-p}$ (horizontal and vertical disks respectively), these
collections are called {\it flow boxes} and they are defined in such
a way that they intersect each $L_\alpha$ in a set of horizontal
disks $\{ {\bf D}^p \times \{y\}\}$. The disks are smoothly
embedded, such that the tangent planes vary continuously on the flow
boxes.

A $(n-p)$ submanifold $T$ of $M$ is called a {\it transversal} if it
is transversal to each $L_\alpha$ of the foliation ${\cal F}_p$. A
{\it transversal Borel measure}  of the foliation ${\cal F}_p$
assigns to each small\footnote{A submanifold is said to be small if
it is contained in a single flow box.} transversal submanifold $T$ a
measure $\mu_{p,T}$. We assume that the measures are holonomy
invariant and they are finite on compact subsets of the transversals
\cite{Plante}.

Thus a {\it geometrical current} is the triple $(L_\alpha,\mu_T,
\nu)$, with the entries being objects defined as above and $\nu$ is
the orientation of $L_\alpha$, which is assigned to every point.

Assume that $M$ is covered by a system of flow boxes endowed with
partitions of unity. Then every $p$-form $\omega$ can be decomposed
into a finite sum $\omega = \sum_i \omega_i$, where each $\omega_i$
has his own support in the $i$-th flow box. We proceed to integrate
out every $\omega_i$ over each horizontal disk $({\bf D}^p \times
\{y\})_i$. Thus we obtain, using the transversal measure,
a continuous function $f_i$ over
$({\bf D}^{n-p})_i$, . In this way we
define a geometric current as
\begin{equation}
\langle (L_\alpha, \mu_T, \nu), \omega \rangle = \sum_i \int_{({\bf
D}^{n-p})_i} \mu_T(y) \left( \int_{({\bf D}^p \times \{y\})_i}
\omega_i \right) .
\end{equation}
This current is closed in the sense of de Rham \cite{dRham} i.e., if
$\omega = d \phi$ where $\phi$ has compact support then $\langle
(L_\alpha, \mu_T, \nu), d\phi \rangle = 0$, since we can write $\phi
= \sum_i \phi_i$. Ruelle and Sullivan \cite{RS} have shown that this
current determines precisely an element of the $p$-th cohomology
group of $M$. It does not depend of the choice of flow boxes. Recall
that any $(n-p)$-form $\rho$ on $M$, determines a $p$-dimensional
current by Poincar\'e duality $\langle \rho, \omega \rangle = \int_M
\rho \wedge \omega$. Thus $(L,\mu,\nu)$ determines an element in
Hom$\big(H^k(M,\mathbb{R}),\mathbb{R}\big)$ which is isomorphic to
$H_k(M,\mathbb{R})$ and therefore gives the asymptotic cycle.

Now consider an example of a geometrical current. Let $\mu_p$ be an
invariant (under ${\bf X}_p$) volume $n$-form and ${\bf X}_p$ is a
$p$-vector field nowhere vanishing on $M$. This defines a current in
the de Rham sense via the $(n-p)$-form $\eta= i_{{\bf X}_p}
(\mu_p)$. The current is given by
\begin{eqnarray}
\label{currentW} W_{\mu, {\bf X}_p} (\beta)  = \int_M i_{{\bf X}_p}
(\mu_p) \wedge \beta.
\end{eqnarray}
This current is not in general closed but it will be closed, for
instance, if the $p$-vector ${\bf X}_p$ consists of vector fields
corresponding to one-parameter subgroups of an action of a Lie group
which preserves the volume form $\mu_p$. More precisely,  one can
obtain asymptotic cycles for values of $p>1$ \cite{SH} as follows.
Consider the action of a connected Lie group $G$ on a smooth compact
oriented manifold $M$, whose orbits are of the same dimension $p$. A
\textit{quantifier} is a continuous field of $p$-vectors on $M$
everywhere tangent to the orbits and invariant under the action of
$G$ via the differential.

A quantifier is said to be \textit{positive} if it is nowhere
vanishing and determines the orientation of the tangent space. A
\textit{preferred action} is an oriented action of a connected Lie
group $G$ such that for any $x \in M$ the isotropy group $I_x$ of
$x$ is a normal subgroup of $G$ and $G/I_x$ is unimodular.

In \cite{SH} it was proved that a preferred action possesses a
positive quantifier and given a positive quantifier we can define a
$1-1$ correspondence between finite invariant measures and
transversal invariant measures. An important result which will be
used in the next sections is the following theorem (Schwartzman
\cite{SH}) that states: Let ${\bf X}_p$ be a positive definite
quantifier (i.e. $p$-vector field) and $\mu_p$ an invariant measure
given by a the volume $n$-form, then $i_{{\bf X}_p} (\mu_p)$ is a
closed $(n-p)$-form and the asymptotic cycle $W_{\mu,{\bf X}_p}$
will be obtained by Poincar\'e duality of an element of $H^{n-p} (M,
\mathbb{R})$ determined by $i_{{\bf X}_p} (\mu_p)$.

If $W_{\mu,{\bf X}_p}$ in $H_p (M,\mathbb{R})$ is an asymptotic
cycle, the theorem gives an explicit way to construct asymptotic
cycles and interpret currents as winding cycles, if the above
conditions are satisfied. In \cite{S} Sullivan gave another way of
specifying a foliation, using structures of $p$-cones and operators
acting over vectors on these cones.

One concrete example of the above is the following: Let $G$ be a
connected abelian Lie group (for instance $\R^n$ or a compact torus
${\mathbb T}^n$) acting differentiably and locally freely (i.e. the
isotropy group of every point is a discrete subgroup of $G$) on the
smooth closed manifold $M$. Then, the orbits of the action
determines a foliation with leaves of the same dimension as $G$.
Since the group is abelian it has an invariant volume form and we
obtain a natural foliated cycle.

\section{Overview of Cohomological Quantum Field Theory: Donaldson-Witten Invariants}

In this section we overview briefly the Donaldson-Witten invariants
for a closed, oriented and Riemannian four-manifold $M$
\cite{Donaldson,DK}, representing our spacetime. We will focus on
the Witten description \cite{WTQ} in terms of correlation functions
(expectation values of some BRST-invariant operators). A
cohomological field theory is a field theory with a BRST-like
operator ${\cal Q}$ transforming as a scalar with respect to the
spacetime symmetries. This operator represents a symmetry of the
theory and it is constructed such that ${\cal Q}^2=0$. The
Lagrangians of these theories can be written as a BRST commutator
(BRST-exact) $L=\{{\cal Q},V \}$, for some functional $V$. Given the
properties of ${\cal Q}$, it implies that the Lagrangian is
invariant under the ${\cal Q}$ symmetry $\{{\cal Q},L\}=0$ i.e. the
Lagrangian is ${\cal Q}$-closed. In general, all observables ${\cal
O}$ in the theory are BRST invariant and they define cohomology
classes given by ${\cal O} \sim {\cal O} + \{ {\cal Q}, \lambda \}$
for some $\lambda$. Here ${\cal O}$ are the observables of the
theory, which are invariant polynomials of the fields under the
symmetry generator ${\cal Q}$. The observables are given by local
field operators thus they depends on the point $x \in M$. Sometimes,
in order to simplify the notation, we will omit explicitly this
dependence.

Usually the relevant topological (twisted) Lagrangian can be derived
from a physical Lagrangian which may depend on the Riemannian metric
$g_{\mu \nu}$ of the underlying spacetime manifold $M$ and
consequently there exists an energy-momentum tensor $T_{\mu \nu}$
which is also BRST-exact, i.e. $T_{\mu \nu}=\{{\cal Q},\lambda_{\mu
\nu}\}$, for some $\lambda_{\mu \nu}$. It was proved for any
BRST-exact operator $\mathcal{O}$ that the correlation function
$\langle \{ \mathcal{Q}, \mathcal{O} \} \rangle$ vanishes and the
partition function is also independent of the metric and the
physical parameters encoded in the Lagrangian. Thus the correlation
functions are topological invariants.

There are several examples of these kind of theories. In particular,
the theories that we are interested in are the Donaldson-Witten and
the Seiberg-Witten ones. From the physical point of view these
theories come from a suitable twist to the Lorentz group of the ${\cal
N}=2$ supersymmetric Yang-Mills theories with a compact Lie group
(for definiteness we will use $SU(2)$ though its generalization to
higher dimensional groups is not difficult). The supercharges also
are affected by the twist which gives rise to our ${\cal Q}$
transforming as a scalar in the new assignation of the
representations of spacetime global symmetries.

In the path integral formalism the Donaldson-Witten polynomials are given by
correlation functions in the Euclidean signature
\begin{equation}
\langle {\cal O} \rangle = \int \mathcal{D}
{\cal X} \exp\big(- L_{DW} / e^2 \big) {\cal O},
\end{equation}
where $L_{DW}$ is the Donaldson-Witten Lagrangian, $e$ is the
coupling constant, the $\mathcal{D} {\cal X}$ represent the measure
of the fields in the theory which includes a non-abelian gauge
$A^a_\mu(x)$, scalar $\phi(x)$, fermionic $\psi(x)$ and ghost
(anti-ghost) fields, all of them taking values in the adjoint
representation of the gauge group. Fields
$(A^a_\mu(x),\psi(x),\phi(x))$ (with associated ghost numbers
$U=(0,1,2)$) constitutes a fermionic BRST multiplet.
We note
that there is a nice mathematical interpretation of the
mentioned ingredients of the theory. For instance, the fields will
represent differential forms on the moduli space of the theory, the
ghost number of the fields corresponds with the degree of these
differential forms and the BRST charge ${\cal Q}$, which changes the
ghost number in a unit, can be regarded as the exterior derivative.

It is possible to see that the change of the correlation functions
with respect to the coupling constant $e$ is ${\cal Q}$-exact. Due
to the property mentioned above $\langle \{{\cal Q},V\}\rangle=0$,
for some $V$, the correlation functions are independent of $e$.
Consequently they can be computed in the semi-classical regime when
$e$ is small and they can be evaluated by the stationary phase
method. The path integration in the space ${\cal C} = {\cal B}/{\cal
G}$ (of all gauge connections ${\cal B}$ modulo gauge
transformations ${\cal G}$) localizes precisely in the space of
gauge fields satisfying the instanton equation (anti-Self-dual
Yang-Mills equations): $\widetilde{F}_{\mu \nu}= - F_{\mu \nu}$ i.e.
the instanton moduli space $\mathcal{M}_D$ of dimension
$d(\mathcal{M}_D ) = 8 p_1 (E) - \frac{3}{2} (\chi + \sigma))$
\cite{AHS} . Here $\chi$ and $\sigma$ are the Euler characteristic
and the signature of $M$ respectively.

It is worth mentioning that in general ${\cal M}_D$ has
singularities which are associated with the reducible connections or
the zero size instantons (small instantons). If one considers four
manifolds with $b_2^+(M)>1$ the moduli space ${\cal M}_D$ behaves as
an smooth, orientable and compact manifold \cite{Donaldson,DK}. But
for general $b_2^+(M)$ it is more involved. Thus, in general, they
are usually neglected by assuming that the only zero modes come from
the gauge connection $A^a_\mu(x)$ and its BRST partner
$\psi_\mu^a(x)$. Scalar field $\phi^a(x)$ has zero modes in the
singularities and this would lead to a modification of the
observables.  The Donaldson map $H_p(M) \to H^{4-p}({\cal M}_D)$
($p=0,\dots ,4$) is given by $\gamma_p \mapsto \int_{\gamma_p}
c_2({\cal P})$, where ${\cal P}$ is the universal bundle over $M
\times {\cal M}_D \subset M \times {\cal B}/{\cal G}$ and $c_2$ is
the second Chern class of ${\cal P}$.

For the gauge group $SU(2)$, the observables are
\begin{equation}
\label{obs}
\mathcal{O}^{\gamma_p} \equiv \int_{\gamma_p} W_{\gamma_p} ,
\end{equation}
where $\gamma_p$ is a $p$-homology cycle of $M$ and  $W_{\gamma_p}$
is a $p$-form over $M$ given by
\begin{eqnarray}
\label{forms}
\label{eq:2}  W_{\gamma_0}(x) = \frac{1}{8 \pi^2} \rm{Tr} \phi^2\, \ \ \ \ \
W_{\gamma_1} = \frac{1}{4 \pi^2} {\rm Tr} \big( \phi
\wedge \psi \big),\nonumber \\
W_{\gamma_2} = \frac{1}{4 \pi^2} {\rm Tr}\big(- i \psi \wedge \psi +
\phi F\big), \ \ \ \ \ W_{\gamma_3} = \frac{1}{4 \pi^2} {\rm
Tr}\big(\psi \wedge F\big), \ \ \ \ \ W_{\gamma_4} = \frac{1}{8
\pi^2} {\rm Tr} \big(F \wedge F\big),
\end{eqnarray}
where $W_{\gamma_0}(x)$ is, by construction, a Lorentz and ${\cal
Q}$ invariant operator. These observables have ghost number
$U=(4,3,2,1,0)$ respectively and they are constructed as descendants
which can be obtained from the relation
\begin{equation}
d W_{\gamma_p} = \{ {\cal Q}, W_{\gamma_{p+1}} \} .
\label{descendent}
\end{equation}
This construction establishes an isomorphism between the BRST
cohomology $H^*_{BRST}({\cal Q})$ and the de Rham cohomology
$H^*_{dR}(M)$. One can check that ${\cal O}^{\gamma_p}$ is
BRST-invariant (BRST-closed)
\begin{equation}
\{ {\cal Q}, \mathcal{O}^{\gamma_p} \} = \int_{\gamma_p} \{ {\cal
Q}, W_{\gamma_p} \} = \int_{\gamma_p} d W_{\gamma_{p-1}} = 0 .
\label{ecuaciondiez}
\end{equation}
For that reason the BRST commutator of $\mathcal{O}^{\gamma_p}$ only
depends of the homology class of $\gamma_p$. Indeed, suppose that
$\gamma_p = \partial \beta_{p+1}$, then we get (BRST-exact)
\begin{equation}
\mathcal{O}^{\gamma_p} = \int_{\gamma_p} W_{\gamma_p} =
\int_{\beta_{p+1}} d W_{\gamma_p} = \int_{\beta_{p+1}} \{ {\cal Q} ,
W_{\gamma_{p+1}} \} = \{ {\cal Q}, \int_{\beta_{p+1}}
W_{\gamma_{p+1}} \} .
\label{ecuaciononce}
\end{equation}
Then the correlation functions are written as
\begin{eqnarray}
 \big\langle \mathcal{O}^{\gamma_{p_1}} \cdots
\mathcal{O}^{\gamma_{p_r}} \big\rangle & = &  \bigg\langle  \prod_{j=1}^r \int_{{\gamma}_{p_j}}
W_{{\gamma}_{p_j}} \bigg\rangle \nonumber \\
& = & \int  \mathcal{D} \mathcal{X} \exp(-L_{DW}/e^2) \prod_{j=1}^r
\int_{\gamma_{p_j}} W_{\gamma_{p_j}} . \label{eq:inv1}
\end{eqnarray}
These are the Donaldson-Witten invariants in the path integral
representation. They are invariants of the smooth structure of $M$.
For simply connected manifolds $\pi_1(M)=0$ the relevant cycles are
of the dimensions $p = 0$, $2$ and  $4$.

Consider a simply connected four-manifold $M$. The correlation
functions of $r$ observables ${\cal O}_{\Sigma_1}, \dots ,{\cal
O}_{\Sigma_r}$ is given by

\begin{eqnarray}
 \big\langle \mathcal{O}_{\Sigma_1}(x_1) \cdots
\mathcal{O}_{\Sigma_r}(x_r) \big\rangle & = &  \bigg\langle  \prod_{j=1}^r
\mathcal{O}_{\Sigma_j}(x_j) \bigg\rangle \nonumber \\
& = & \int  \mathcal{D} \mathcal{X}
\exp(-L_{DW}/e^2) \prod_{j=1}^r \mathcal{O}_{\Sigma_j}(x_j),
\label{eq:in}
\end{eqnarray}
where we have considered only $r$ arbitrary 2-cycles  i.e.
$\gamma_{2_j}=\Sigma_j$ with $j=1,\dots,r$. Since
$\mathcal{O}_{\Sigma_j}(x_j)$ has ghost number $U=2$, the above
correlation function has $U=2r$. In terms of the zero modes one can
write each $\mathcal{O}_{\Sigma_j}(x_j)= \Phi_{i_1 i_2}(a_i)
\psi^{i_1} \psi^{i_2}$, which absorbs two zero modes and
consequently in the weak coupling limit we have
\begin{equation}
\big\langle \mathcal{O}_{\Sigma_1}(x_1) \cdots
\mathcal{O}_{\Sigma_r}(x_r) \big\rangle  =  \int_{{\cal M}_D}  \Phi_{\Sigma_1} \wedge \cdots \wedge \Phi_{\Sigma_r}.
\label{mapsso}
\end{equation}
Thus we have $\Sigma \in H_2(M) \to \Phi_\Sigma \in H^2({\cal M}_D)$
and Eq.(\ref{mapsso}) becomes:
\begin{equation}
\big\langle \mathcal{O}_{\Sigma_1}(x_1) \cdots
\mathcal{O}_{\Sigma_r}(x_r) \big\rangle  = \#(H_{\Sigma_1} \cap
\cdots \cap H_{\Sigma_r}),
\label{intersectionNo}
\end{equation}
where $H_{\Sigma_j}$ is the Poincar\'e dual to $\Phi_{\Sigma_j}$ and
represents a $(d({\cal M}_D)-2)$-homology cycle of the instanton
moduli space ${\cal M}_D$. Equation (\ref{intersectionNo}) is
interpreted as the intersection number of these homology cycles in
the moduli space.

The topological invariance is not evident from the Eq.
(\ref{eq:inv1}). However the above construction has a natural
interpretation in terms of equivariant cohomology
\cite{Kanno:1988wm}. Moreover Atiyah and Jeffrey \cite{AJ} showed
that this expression can be reinterpreted in terms of the Euler
class of a suitable infinite dimensional vector bundle in the
Mathai-Quillen formalism \cite{MQ}. This construction requires a
real vector bundle ${\cal E}$ over the quotient space ${\cal C}$ of
the space of all connections ${\cal B}$ modulo gauge transformations
${\cal G}$. This bundle is such that the fibres are the space of
sections $\Gamma(\Lambda^{2,+} \otimes {\rm ad}(P))$. Here $P$ is
the SU(2)-principal bundle over $M$ with gauge connection
$A^a_\mu(x)$.

Moreover a section $s$ of ${\cal E}$ is given by  $s=-F^+$, i.e. the
locus $s^{-1}(0)$ is precisely the anti-self-dual moduli space
${\cal M}_D \subset {\cal C}$. The Euler class $e({\cal E})$ is the
pullback $s^*\Phi({\cal E})$ of the Thom class $\Phi({\cal E})$ of
${\cal E}$, under the section $s$.  If $d({\cal C})$ is the
dimension of ${\cal C}$, the work of Mathai-Quillen \cite{MQ} allows
to gave a gaussian representative for the associated Thom class
given by $e_{s,\nabla} ({\cal E}) = \exp [ - {1 \over e^2} |s|^2 +
\cdots],$ (which is given by a $2m$ differentiable form on ${\cal
C}$) such that the Euler class is given by
\begin{equation}
\int_{ {\cal C}} e_{s,\nabla} ({\cal E}) \wedge \alpha,
\label{eulerc}
\end{equation}
where $\alpha$ is an appropriate differential form of co-dimension
$2m$ i.e. $\alpha \in H^{d({\cal C})-2m}({\cal C})$. This Euler
class is of course independent on the connection $\nabla$ used in
this construction. Euler class (\ref{eulerc}), for an appropriate
$\alpha$, represents the Donaldson-Witten invariants
(\ref{eq:inv1}).

\vskip 2truecm
\section{Donaldson-Witten Invariants for Flows}

In this section we study the Donaldson-Witten invariants when there
exist flows associated to a $p$-vector field over the spacetime
manifold $M$ equipped with an invariant probability measure $\mu$
(normalized such that $\int_M \mu =1$) and a non-singular $p$-vector
field ${\bf Y}_p= Y_1 \wedge \cdots \wedge Y_p$ where $Y_i=Y^\mu_i
\partial_\mu$ with $\mu=1,\dots,4$ and $i=1, \cdots , p$. We
require that the probability measure $\mu_{T,p}$ be invariant under
${\bf Y}_p$ for every $p$.  Thus, the global information is encoded
in the set of triplets $\{(M,{\cal F}_p,\mu_{T,p}), \
p=0,\dots,4\}$, where ${\cal F}_p$ is the foliation generated by the
$p$-vector field ${\bf Y}_p$. Each triplet $(M,{\cal
F}_p,\mu_{T,p})$ determines an asymptotic $p$-cycle that we denote
as $\widetilde{\gamma}_p$.

Now we define the generalized Lie derivative ${\cal L}_{{\bf Y}_p}$
for $p$-vectors which is a graded operator defined as follows
\begin{equation}
{\cal L}_{{\bf Y}_p} \omega = [d,i_{{\bf Y}_p}] \omega  = d (i_{{\bf Y}_p} \omega ) + (-1)^{\deg d \cdot \deg {i_Y}} i_{{\bf Y}_p} (d \omega),
\end{equation}
being $[d,i_{{\bf Y}_p}]$ a graded commutator and $\omega$ is a
$p$-form.  For further details on the formalism of multi-vector
field see \cite{vaisman,Holm}. Here there are two possibilities:

\begin{itemize}
\item{} The homology groups associated to the orbits of the $p$-vector fields ${\bf Y}_p$'s are
trivial therefore they will not give relevant information of the
four-manifold, however we can use these trivial asymptotic homology
cycles to describe some particular interesting configurations of
flows. This case corresponds to the situations found in Refs.
\cite{VV,GarciaCompean:2009zh}. In particular if $\omega$ is the
volume form invariant under ${\bf Y}_p$ for every $p$, the last term
of the previous equation is zero, then the Lie derivative ${\cal
L}_{{\bf Y}_p} \mu_p = d (i_{{\bf Y}_p} \mu_p )$, the term $i_{{\bf
Y}_p} \mu_p$ looks like the expression from the Schwartzman theorem
at the end of Sec. 2. In order for $i_{{\bf Y}_p} \mu_p$ to be a
cohomology class it needs to be closed. This requirement is
established by the following equation
\begin{equation}
\label{freediv} {\cal L}_{{\bf Y}_p} \mu_p =0.
\end{equation}
If this condition implies $d (i_{{\bf Y}_p} \mu_p) =0$ i.e. $i_{{\bf
Y}_p} \mu_p$ is closed then this element defines an element of the
$(n-p)$ cohomology group.

\item{} The other possibility corresponds to the case when the $\widetilde{\gamma}_p$ are orbits of the flow
generated by the $p$-vector fields  ${\bf Y}_p$ for each value of
$p$. These cycles are non-trivial. In this case it gives rise to a
generalization of the four-manifold invariants as the measure $\mu$
is supported in the whole manifold $M$.

\end{itemize}
In the present paper we will focus mainly on this second possibility.

\subsection{The Definition of Observables}

Now we will introduce flows over $M$, and promote  its homology
cycles to asymptotic cycles. We define the asymptotic observable for a $p$-vector
fields ($p=1, \ldots, 4$)  according to the expression

\begin{equation}
\widetilde{\cal O}_{{\bf Y}_p}(\mu_p) = \int_{\widetilde{\gamma}_p}
W_{\gamma_p}:= \int_M  i_{{\bf Y}_p}(W_{\gamma_p})\mu_p(x) ,
\end{equation}
where $\mu_p$ is the volume form of $M$ invariant under ${\bf Y}_p$
and $i_{{\bf Y}_p}(W_{\gamma_p})$ denotes the contraction and  ${\rm
Tr}$ is the trace of the adjoint representation of the gauge group.

We can think of the observable as an average winding number of the
asymptotic cycle. The observables are related to the asymptotic
cycles then they carry information about the flow whether it is
trivial or not.

Let ${\bf Y}_1 ,\ldots , {\bf Y}_4$ be $p$-vector fields ($p=1,
\ldots , 4$) and together with the expressions $(\ref{obs})$ and
$(\ref{forms})$ we define the asymptotic observable as

\begin{eqnarray}
\label{eq:f1} \widetilde{\mathcal{O}}_{ {\bf Y}_0}(\mu_0) \equiv
{\cal O}^{\gamma_0}(x)= {1\over 8 \pi^2}{\rm Tr}\phi^2,\\
\label{onett} \widetilde{\mathcal{O}}_{ {\bf Y}_1}(\mu_1) =  \int_M
{\rm Tr} \frac{1}{4 \pi^2} i_{{\bf Y}_1}(\phi \psi)  \,\mu_1,\\
\label{eq:f2} \widetilde{\mathcal{O}}_{{\bf Y}_2}(\mu_2) = \int_M
{\rm Tr} \frac{1}{4 \pi^2} i_{{\bf Y}_2}(-i \psi \wedge \psi + \phi
F) \, \mu_2,\\
\label{eq:f3} \widetilde{\mathcal{O}}_{{\bf Y}_3}(\mu_3) = \int_M
{\rm Tr} \frac{1}{4 \pi^2} i_{{\bf Y}_3} (\psi \wedge F) \, \mu_3,\\
\label{eq:f4} \widetilde{\mathcal{O}}_{{\bf Y}_4}(\mu_4) = \int_M
{\rm Tr} \frac{1}{8 \pi^2} i_{{\bf Y}_4}{(F \wedge F)}
 \, \mu_4.
\end{eqnarray}

It is an easy matter to check that these asymptotic observables
$\widetilde{\mathcal{O}}_{{\bf Y}_p}(\mu_p)$ are ${\cal
Q}$-invariant

\begin{eqnarray}
\{ {\cal Q} , \widetilde{\mathcal{O}}_{ {\bf Y}_{p+1} }(\mu_{p+1})
\}& = &\{ {\cal Q} , \int_M i_{{\bf Y}_{p+1}}(W_{\gamma_{p+1}})
\mu_{p+1} \} \nonumber
\\ & = &
\int_M \{{\cal Q}, i_{{\bf Y}_{p+1}} (W_{\gamma_{p+1}}) \}  \mu_{p+1} \nonumber \\
 & = &
\int_M i_{{\bf Y}_{p+1}} (d W_{\gamma_p}) \mu_{p+1} = 0 .
\label{veinticinco}
\end{eqnarray}
Here we have used the fact that the measure is invariant under the
flow i.e. these observables are closed in the de Rham sense (see
theorem 2A from Ref. \cite{SH}) and the fact that the BRST charge
${\cal Q}$ commutes with the contraction operation $i_{{\bf Y}_p}$.
Then these asymptotic observables are BRST invariant, therefore they
will give rise to topological invariants of dynamical system through
a generalization of the Donaldson-Witten invariants.

Moreover we observe that the problem arising in the Jones-Witten
case \cite{VV}, which distinguishes strongly the abelian and
non-abelian cases is absent here and for the present case, the
non-abelian case can be treated exhaustively. Even if the theory is
non-abelian, our observables are Lie algebra-valued $p$-forms and
the group and space-time information decouples. For the gauge group
SU$(N)$ with Lie algebra su$(N)$ we take for instance
\begin{eqnarray}
\widetilde{\mathcal{O}}_{{\bf Y}_2}(\mu_2) = \int_{M} \frac{1}{4
\pi^2} {\rm Tr} i_{{\bf Y}_2} (-i\psi \wedge \psi + \phi F) \  \mu_2
\nonumber \\
= \int_{M} \frac{1}{4 \pi^2} {\rm Tr} \{ i_{{\bf Y}_2}(-i \psi^a
\wedge \psi^b + \phi^a F^b ) t_a t_b \} \  \mu_2 ,
\end{eqnarray}
where if $t_a$ and $t_b$ are generators of su$(2)$ they satisfy the
normalization condition: ${\rm Tr}( t_a t_b ) = {1\over 2} \delta_{a
b}$. Then last expression takes the following form
\begin{equation}
\widetilde{\mathcal{O}}_{{\bf Y}_2}(\mu_2) = \int_{M} \frac{1}{8
\pi^2} i_{{\bf Y}_2} (-i \psi^a \wedge \psi^a + \phi^a F^a ) \
\mu_2 .
\end{equation}
We will use the following notation for $r$ components of $p$-cycles
of different dimension. The observables will be denoted by
$\widetilde{\mathcal{O}}_{{\bf Y}_{p_j}} (\mu_{p_j})$, where $p_j$
take values $1, \ldots,4$, $j=1,\dots ,r$ and such that they satisfy
$\sum_{p,j} p_j = \rm{d} (\mathcal{M}_D)$, which is the dimension of
the moduli space of instantons.

\subsection{Donaldson-Witten Invariants of four-manifolds for Flows}

For an oriented manifold $M$ with $p_j$-vectors fields ${\bf
Y}_{p_j}$, with probability invariant measure $\mu_{p_j}$, the
$r$-point correlation functions (Donaldson-Witten invariants) for
flows ${\bf Y}_{p_j}$ is given by

\begin{eqnarray}
\bigg\langle\widetilde{\mathcal{O}}_{{\bf Y}_{p_1}} (\mu_{p_1})
\cdots \widetilde{\mathcal{O}}_{{\bf Y}_{p_r}}
(\mu_{p_r})\bigg\rangle = \int \mathcal{D} \mathcal{X} \exp (-
L_{DW}/e^2 ) \prod_{j=1}^r \int_M \,
i_{{\bf Y}_{p_j}}(W_{\gamma_{p_j}}) \mu_{p_j}(x) . \nonumber \\
\label{eccf}
\end{eqnarray}
This expression is reduced to the ordinary Donaldson-Witten
invariants $(\ref{eq:inv1})$, when the measure is supported on the
cycles. One can think of this set of measures $\{\mu_{p_j}\}$ as
Dirac measures on the set of $p-j$-cycles  $\{
\widetilde{\gamma}_{p_j}\}$. If we consider the invariant
probability measure $\mu_p = \sum_{j}^r \mu_{p_j}$, where each
$\mu_{p_j}$ is supported on $\widetilde{\gamma}_{p_j}$ and they are
uniformly distributed with respect the coordinates of $\{
\widetilde{\gamma}_{p_j}\}$. In other words $\mu_{p_j}$ is supported
on $\gamma_{p_j}$ and it coincides with the normalized area form of
the surface $\gamma_{p_j}$. We need to normalize in order to have
$\mu$ a probability measure.

We want to remark that the underlying $p_j$-fields ${\bf Y}_{p_j}$
will be considered here just as spectator fields. That is, they are
background fields that are not of dynamical nature and don't
represent additional degrees of freedom of the underlying theory.
Thus, they don't contribute to the measure $\mathcal{D}
\mathcal{X}$, to the Lagrangian $L_{DW}$ nor to the counting of zero
modes and consequently they do not lead to a modification of the
dimension of the moduli space of instantons\footnote{Thus we compute
the effect of these spectator vector fields on the invariants. At
this stage it is not possible  to compute the back reaction of all
dynamical fields to ${\bf Y}_{p_j}$.}. There will be an influence of
these $p_j$-vector fields to our systems modifying mainly the
structure of the observables of the theory. The structure of the
vacuum also remains unchanged i.e. the mass gap and the chiral
symmetry breaking are still playing an important role in the
definition of invariants.

For the moment we will consider an arbitrary operator
$\widetilde{\mathcal{O}}_{{\bf Y}_{p_j}} (\mu_{p_j})$ with $p_j \geq
1$ (because in the case $p_j=0$ there is not a flow).  Now we
proceed to perform the integral over the non-zero modes, as in the
case without flows.

We assume that the only zero modes correspond to the gauge field
$A_\mu$ and those associated to $\psi_\mu$. Denote this observable
by $\widetilde{\mathcal{O}}_{{\bf Y}_{p_j}} (\mu_{p_j}) =
\widetilde\Phi_{i_1 \cdots  i_n}(a_i,{\bf Y}_{p_j}) \psi^{i_1}
\ldots \psi^{i_n}$, where $a$'s denotes the zero modes of the gauge
field and $\psi$'s are the zero modes of the fermionic field,
$\widetilde\Phi (a_i, {\bf Y}_{p_j})$ is a function that only
depends on the zero modes of the gauge field and contains the
information about the flow. As in the standard case the partition
function is zero, the integrals which are non-zero are of the form
$(\ref{eccf})$, where $\widetilde{\mathcal{O}}$ absorb the zero
modes.

Performing the functional integration over the non-zero modes in the
weak coupling limit we get $\widetilde\Phi_{i_1 \cdots i_n}(a_i,{\bf
Y}_{p_j})$ is an skew-symmetric tensor then $\mathcal{\widetilde O}$
can be regarded as a $n=d({\cal M}_D)$-form in $\mathcal{M}_D$.
Consequently the correlation functions of one observable
$\widetilde{\cal O}$ reads
\begin{eqnarray}
\bigg\langle \widetilde{\mathcal{O}}_{{\bf Y}_{p_j}}
(\mu_{p_j})\bigg\rangle & = & \int_{\mathcal{M}_D} da_1 \ldots da_n
d\psi^1 \ldots d\psi^n \widetilde{\Phi}_{i_1 \cdots  i_n}(a_i,{\bf
Y}_{p_j}) \psi^{i_1} \ldots \psi^{i_n} \nonumber \\& = &
\int_{\mathcal{M}_D} \widetilde{\Phi}_{{\bf Y}_{d({\cal M}_D)}},
\end{eqnarray}
where we integrate out the $a_i$'s and obtain a $n$-form
$\widetilde{\Phi}$ defined in the moduli space ${\cal M}_D$. If one
considers a product of observables $\widetilde{\mathcal{O}} =
\widetilde{\mathcal{O}}_{{\bf Y}_{p_1}} (\mu_{p_1}) \cdots
\widetilde{\mathcal{O}}_{{\bf Y}_{p_r}} (\mu_{p_r})$ with
$\sum_{p,j} p_j = n= d({\cal M}_D)$ and $p_j$ being the number of
zero modes of $\widetilde{\mathcal{O}}_{{\bf Y}_{p_j}} (\mu_{p_j})$
then, in analogy to Ref. \cite{GarciaCompean:2009zh} one obtains:
\begin{equation}
 \bigg\langle\widetilde{\mathcal{O}}_{{\bf Y}_{p_1}}
(\mu_{p_1}) \cdots \widetilde{\mathcal{O}}_{{\bf Y}_{p_r}}
(\mu_{p_r})\bigg\rangle = \int_{{\cal M}_D} \widetilde{\Phi}_{{\bf
Y}_{p_1}} \wedge \cdots \wedge \widetilde{\Phi}_{{\bf Y}_{p_r}}.
\label{integralmoduli}
\end{equation}
These correlation functions are the asymptotic Donaldson-Witten
invariants and they are invariants of the triplet $(M,\mathcal F,
\mu)$. In order to compute the observables we integrate out the zero
modes. This is completely analogous to the case without flows
because the measure of the path integral does not include the ${\bf
Y}'s$
\begin{eqnarray}
\label{eq:fnew} \widetilde{\Phi}_{{\bf Y}_{0}} = \frac{1}{8 \pi^2}
{\rm Tr} \langle \phi \rangle^2,\\ \label{eq:1p}
\widetilde{\Phi}_{{\bf Y}_{1}} = \int_M {\rm Tr} \frac{1}{4 \pi^2}
i_{{\bf Y}_1}(\langle \phi \rangle \psi) \mu_1,\\ \label{eq:f2p}
 \widetilde{\Phi}_{{\bf Y}_{2}} = \int_M {\rm Tr} \frac{1}{4 \pi^2} i_{{\bf Y}_2}(-i \psi
\wedge \psi + \langle \phi \rangle F) \mu_2,\\ \label{eq:f3p}
 \widetilde{\Phi}_{{\bf Y}_{3}} = \int_M {\rm Tr} \frac{1}{4 \pi^2}i_{{\bf Y}_3}(\psi \wedge
F) \mu_3,\\ \label{eq:f4p}
 \widetilde{\Phi}_{{\bf Y}_{4}} = \int_M {\rm Tr} \frac{1}{8 \pi^2} i_{{\bf Y}_4} (F \wedge F)
 \mu_4.
\end{eqnarray}
Now we define the intersection number in a way analogous to the case without flows.

For the simply connected case $(\pi_1 (M) = 0)$, we have that the
important observables are those associated with cycles of dimension
$0$, $2$ and $4$\footnote{0 and 4 cycles are related by Hodge
duality, so we will consider one of them, say 0-cycles.}. In general
a $p_j$-cycle has associated an operator (form) with ghost number
$U= 4 - p_j$, corresponding to the Donaldson map $\mu_D :H_p (M) \to
H^{4-p} (\mathcal{M}_D)$. In Ref. \cite{WTQ} Witten constructed this
map, interpreted as intersection number of cycles in the four
manifold $M$
\begin{eqnarray}
\label{donaldsoninvs} \nonumber \bigg\langle\widetilde{I}_{{\bf
Y}_{2_1}} (\mu_{2_1})(x_1) \cdots \widetilde{I}_{{\bf Y}_{2_r}}
(\mu_{2_r})(x_r)\bigg\rangle & = & \int_{{\cal M}_D} \nu_{{\bf Y}_{2_1}} \wedge \cdots \wedge \nu_{{\bf Y}_{2_r}} \\
& = & \# \big( H_{\widetilde{\Sigma}_{{\bf Y}_{2_1}}}  \cap \dots
\cap H_{\widetilde{\Sigma}_{{\bf Y}_{2_r}}} \big) ,
\end{eqnarray}
where $H_{\widetilde{\Sigma}_{{\bf Y}_{2_r}}}$ is the Poincar\'e
dual of codimension 2. If the observables $\widetilde{I}_{{\bf
Y}_{2_j}}$ are denoted by $\widetilde{I}_{{\bf Y}_{j}}$ and
$\mu_{2_j}$ is denoted by $\mu_{j}$, then the equation
($\ref{donaldsoninvs}$) represents the asymptotic intersection
linking numbers of the $2$-flows in the moduli space ${\cal M}_D$
determined by the integration of differential two-forms $\nu_j$'s on
${\cal M}_D$ depending on the set of 2-vector fields $\{{\bf
Y}_j\}_{j=1, \dots ,r}$ with $r=d/2$. In terms of the asymptotic
cycles Eq. ($\ref{donaldsoninvs}$) represents the asymptotic
intersection number of $r$ asymptotic homology 2-cycles
$\widetilde{\Sigma}_{ {\bf Y}_{j}}$ in $M$.

Donaldson-Witten invariants (\ref{eq:inv1}) are defined for
$b^+_2(M) >1$. It is very interesting to know what is the analog
condition for defining the existence of the corresponding asymptotic
invariants for foliations. The analog of the wall-crossing that does
exist in the Donaldson case for $b_2^+(M)=1$, will be also of
interest in the context of foliations. We leave this question for
future work.

\subsection{Asymptotic Intersection Numbers}

In this subsection we use the dynamics of strongly coupled
supersymmetric gauge theories. In particular we use some features
as: the existence of a mass gap, the cluster decomposition and a
structure of the vacua degeneracy consisting of a finite number of
discrete states obtained after a chiral symmetry breaking due to
gaugino condensation. We proceed to find an interpretation of the
intersection number for asymptotic cycles described in Eq.
(\ref{donaldsoninvs}) with the aid of the mentioned features. In
order to do that we are going to compute the 2-point correlation
function of a pair of the observables $\widetilde{I}_{{\bf Y}_{1}}
(\mu)$ at different points
\begin{equation}
\bigg\langle\widetilde{I}_{{\bf Y}_{1}}
(\mu_1)(x_1)\widetilde{I}_{{\bf Y}_{2}} (\mu_2)(x_2)\bigg\rangle =
\# \big( H_{\widetilde{\Sigma}_{{\bf Y}_1}} \cap
H_{\widetilde{\Sigma}_{{\bf Y}_2}}\big), \label{intersection}
\end{equation}
where the $x_i$'s are points on $M$ and $\# \big(
H_{\widetilde{\Sigma}_{{\bf Y}_1}} \cap H_{\widetilde{\Sigma}_{{\bf
Y}_2}}\big)$ represents the asymptotic intersection number of the
asymptotic cycles $ \widetilde{\Sigma}_{{\bf Y}_1}$ and $
\widetilde{\Sigma}_{{\bf Y}_2}$.

As in the standard case without flows  the one point correlation
function $\big\langle\widetilde{I}_{{\bf Y}} (\mu)(x)\big\rangle$ is
also zero, for the same reason that in the standard case. In
$M=\IR^4$ it vanishes by Lorentz invariance with the measure
invariant under the flow, this yields
$$
\big\langle\widetilde{I}_{{\bf Y}} (\mu)(x)\big\rangle =
\int_{\widetilde{\Sigma}} d\sigma^{mn} \langle Z_{mn}  \rangle
$$
\begin{equation}
= \int_M \langle i_{\bf Y}(Z) \rangle \mu = \int_M \langle
Y^{mn}Z_{mn} \rangle \mu.
\end{equation}
As the ${\bf Y}'$s are not dynamical fields their expectation values
is given by $Y^{mn}\langle Z_{mn} \rangle$ and as $\langle Z_{mn}
\rangle$ vanishes by Lorentz invariance then consequently
$\big\langle\widetilde{I}_{{\bf Y}}(\mu)(x)\big\rangle$ also
vanishes in flat spacetime. However for a general four-manifold in a
theory with a mass gap (it is known that ${\cal N}=2$ gauge field
theories in four dimensions don't have a mass gap, however we
assume, following Witten \cite{MINIMAL}, that is indeed the
case\footnote{In the process of obtaining invariants of smooth
manifolds from physical theories, the dynamics of these theories is
an important guide. However the topological invariants are
independent on the metric and the coupling constant one can compute
these invariants in the limit where the theory is under control. The
assumption of a mass gap for ${\cal N}=2$ theories is justified  as
it allows to compute invariants for some four-manifolds. However it
is observed that the own theory tell us that one has to consider the
full dynamics (including the supersymmetry breaking) in order to
find the right invariants.}), the expectation value of the operator
$Z_{mn} Y^{mn}$ is expanded, as in \cite{MINIMAL} in terms of local
invariants of the Riemannian geometry of $M$
\begin{equation}
\big\langle Z_{mn}(x) Y^{mn}(x)\big\rangle = D_m R D_n D_s D^s R \cdot Y^{mn} \pm \cdots
\end{equation}
Under the metric scaling $g \to tg$ with $t$ positive, the volume
form $\mu$ scales as $t^4$, then  $\big\langle Z_{mn}(x)
Y^{mn}(x)\big\rangle$ should scale faster than $1/t^4$. This is
precisely achieved by the mass in the case we have flows. Thus, in
general $\big\langle \int_{M} i_{\bf Y}(Z)  \, \mu \big\rangle$
vanishes as $t \to \infty$.

Now we want to compute
\begin{equation}
\bigg\langle\widetilde{I}_{{\bf Y}_{1}}
(\mu_1)(x_1)\widetilde{I}_{{\bf Y}_{2}} (\mu_2)(x_2)\bigg\rangle =
\int_{{M}_1 \times {M}_2} G_{{\bf Y}_1 , {\bf Y}_2}(x_1,x_2) \
\mu_1(x_1) \ \mu_2(x_2), \label{internumb}
\end{equation}
where
\begin{equation}
G_{{\bf Y}_1 , {\bf Y}_2}(x_1,x_2) =  \big\langle i_{{\bf Y}_1}(Z)(x_1) \cdot i_{{\bf Y}_2}(Z)(x_2) \big\rangle .
\end{equation}
Considering the properties of $i_{X_1 \wedge \cdots \wedge X_p}$,
one can see that the next formula holds
\begin{equation}
\label{derivation} i_{X_1 \wedge \cdots \wedge X_p} B_p \wedge \mu_n
- (-1)^{{p \over 2}(3+p)} B_p \wedge i_{X_p \wedge \cdots \wedge
X_1}\mu_n=0,
\end{equation}
where $B_p$ is any $p$-form. After some work it is easy to see that
using the previous equation we have
\begin{equation}
\bigg\langle\widetilde{I}_{{\bf Y}_{1}}
(\mu_1)(x_1)\widetilde{I}_{{\bf Y}_{2}} (\mu_2)(x_2)\bigg\rangle =
\int_{M_1 \times M_2} (\Theta_{{\bf Y}_1} \wedge Z)(x_1) \wedge
(\Theta_{{\bf Y}_2} \wedge Z)(x_2) \cdot \delta (x_1-x_2)
\end{equation}
where $\Theta_{{\bf Y}_1}=i_{{\bf Y}_1}(\mu_1)$ and $\Theta_{{\bf
Y}_2}=i_{{\bf Y}_2}(\mu_2)$ are the Poincar\'e dual to
$\widetilde{\Sigma}_1$ and $\widetilde{\Sigma}_2$ respectively. This
is a double form \cite{dRham} and consequently $G_{{\bf Y}_1 , {\bf
Y}_2}(x_1,x_2)$ is proportional to $\delta(x_1-x_2)$ and the only
non-vanishing contributions come from the points $x_1=x_2$.

Equivalently one can follow a dimensional analysis with $g \to tg$,
with $t \to \infty$. For $x_1 \not= x_2$, $\big\langle i_{{\bf Y}_1}
(Z (x_1))\, i_{{\bf Y}_2} (Z (x_2)) \big\rangle$ vanishes faster
than $1/t^8$. The only possible non-vanishing contribution is
localized around $x_1=x_2$ as $t\to \infty$. These are precisely the
intersection points of the asymptotic cycles. That is reduced to the
transversely intersection of the flows in finitely many points. Thus
we have
\begin{equation}
\bigg\langle\widetilde{I}_{{\bf Y}_{1}}
(\mu_1)(x_1)\widetilde{I}_{{\bf Y}_{2}} (\mu_2)(x_2)\bigg\rangle =
\eta \cdot \# \big(\widetilde{\Sigma}_{{\bf Y}_1} \cap
\widetilde{\Sigma}_{{\bf Y}_2}) \cdot \langle 1 \rangle ,
\label{intersecdos}
\end{equation}
where $\eta$ is a constant, $\langle 1 \rangle = \exp(a\chi(M) +
b\sigma(M))$ with $a,b$ being constants and $\chi(M)$ and
$\sigma(M)$ the Euler characteristic and signature of $M$
respectively. In analogy with the definition of asymptotic linking
number we define the asymptotic intersection number of two $2$-flows
generated by the 2-vector fields ${\bf Y}_1$ and ${\bf Y}_2$. Thus
$\big\langle\widetilde{I}_{{\bf Y}_{1}}
(\mu_1)(x_1)\widetilde{I}_{{\bf Y}_{2}} (\mu_2)(x_2)\big\rangle$ can
be interpreted as the average intersection number.

Let $\{ \widetilde{\Sigma}_{{\bf Y}_1}, \dots
,\widetilde{\Sigma}_{{\bf Y}_r}\}$ be a set of $r$ arbitrary
asymptotic homology 2-cycles. With the aid of cluster decomposition
property for a vacua consisting of only one state, Eq.
(\ref{intersecdos}) can be used to write the generating functional
of the correlation functions of observables associated to $r$
2-cycles
\begin{equation}
\bigg\langle\ \exp \bigg( \sum_a \alpha_a \widetilde{I}_{{\bf
Y}_{a}} (\mu_a)\bigg) \bigg\rangle =  \exp\bigg({\eta \over 2}
\sum_{a,b} \alpha_a \alpha_b \# \big(\widetilde{\Sigma}_{{\bf Y}_a}
\cap \widetilde{\Sigma}_{{\bf Y}_b})    \bigg) \cdot \langle 1
\rangle.
\end{equation}
This is given in terms of the pairwise intersection between the
corresponding asymptotic cycles. If one incorporates the operators
$\widetilde{\cal O}$ and take into account that the vacua consist of
a finite set ${\cal S}$ of discrete states it can be modified as
follows
\begin{equation}
\bigg\langle\ \exp \bigg( \sum_a \alpha_a \widetilde{I}_{{\bf
Y}_{a}} (\mu_a) + \lambda \widetilde{\cal O}\bigg) \bigg\rangle =
\sum_{\rho\in {\cal S}} C_\rho\exp\bigg({\eta_\rho \over 2}
\sum_{a,b} \alpha_a \alpha_b \# \big(\widetilde{\Sigma}_{{\bf Y}_a}
\cap \widetilde{\Sigma}_{{\bf Y}_b}) + \lambda \langle {\cal O}
\rangle_\rho \bigg),
\end{equation}
where $C_\rho$ is a constant including the gravitational
contribution of the curvature invariants $\chi(M)$ and $\sigma(M)$
coming from $<1>$. When the gauge group is SU(2) the chiral symmetry
breaking tells that the set ${\cal S}$ is precisely $\IZ_2$. Thus
the above formula consists of two terms.

\section{Donaldson-Witten Invariants for K\"{a}hler Manifolds with
Flows}

In this section we discuss the Donaldson invariants on a K\"ahler
$4$-manifold $M$. We follow closely Ref. \cite{MINIMAL}, from where
we take the notation and conventions. We use the dynamics of strong
coupling ${\cal N}=1$ supersymmetric gauge theories in four
dimensions in the infrared. In particular, the perturbation of the
${\cal N}=2$ theory by adding a mass term\footnote{The mass term
consist of a quadratic term of a scalar superfield in the adjoint
representation of the gauge group.} breaks supersymmetry leaving a
theory with a remnant ${\cal N}=1$ supersymmetry. As we mentioned
before, the properties of strong coupled gauge theories
(confinement, mass gap and chiral symmetry breaking) are an
important subject in order to compute the invariants of K\"ahler
manifolds admitting a non-trivial canonical class in
$H^{(2,0)}(M)\not=0$ \footnote{The condition of the existence of a
canonical class is related to $b_2^+(M)$ by $b_2^+(M)= 2 {\rm
dim}H^{(2,0)}(M)+1$. Thus, the condition $H^{(2,0)}(M)\not=0$ is
equivalent to the familiar one $b_2^+(M)>1$.}.  We consider in
addition a series of non-singular smooth flows generated by 2-vector
fields ${\bf X}$ and ${\bf Y}$ over $M$. We want to describe how
Donaldson-Witten invariants of K\"ahler manifolds will be modified
in the presence of these flows\footnote{Remember that the 2-vector
fields ${\bf X}$ and ${\bf Y}$ are not dynamical and they do not
contribute to the Feynman integral to compute correlation functions.
The analysis of zero modes is also unchanged and the dimension of
the moduli space of instantons remains the same. The only change
will be reflected in the definition of the observables. }.

The theory on $\IR^4$ in euclidean coordinates $(y^1,\dots,y^4)$
with $z_1=y^1+iy^2$ and $z_2=y^3+iy^4$, suggests that the theory
written in terms of ${\cal N}=1$ multiplets implies that the
observables $Z$ are given by
\begin{equation}
Z^{(2,0)}= \psi \psi + \overline{\omega} \overline{B}B, \  \  \ \
Z^{(1,1)}= \lambda \psi + BF, \  \  \ \
Z^{(0,2)}= \lambda \lambda,
\label{observa}
\end{equation}
where $\overline{\omega}$ is an anti-holomorphic 2-form. These
observables are BRST-invariant with respect to the remnant
supercharge $Q_1$ (after the supersymmetry breaking). This structure
of observables comes from decomposition of a ${\cal N}=2$ vector
multiplet in terms of ${\cal N}=1$ gauge multiplet $(A_m,\lambda)$
and a complex matter multiplet given by the scalar superfield
$\Phi=(B,\psi).$ Here $A_m$ is a gauge field, $B$ is a complex
scalar field and $\lambda$ and $\psi$ are spinor fields, all of them
in the adjoint representation of the gauge group. The observables
$Z^{(1,1)}$ and $Z^{(0,2)}$ come from the mentioned decomposition.
However the presence of the term $\overline{\omega} \overline{B}B$
in $Z^{(2,0)}$ is a direct manifestation of the mass term that is
added to the ${\cal N}=2$ Lagrangian in order to break
supersymmetry.  The introduction of a mass term in $\IR^4$ reads
\begin{equation}
\Delta L = - m \int d^4x d^2 \theta {\rm Tr} \Phi^2 - \ {\rm h.c.},
\label{massone}
\end{equation}
where the volume form is $d^4 x d^2 \theta = d^2 z d^2 \overline{z}
d^2 \theta$. This term preserves only $\mathcal{N}=1$ supersymmetry.
It was proved in \cite{MINIMAL} that this perturbation $\Delta L$ to
the Lagrangian is of the form:  $\sum_a \alpha_a I(\Sigma_a) + \{
Q_1, \cdot \}$.  The canonical divisor $C \subset M$ is defined as
the zero locus of $\omega$. Thus, in general on a curved K\"ahler
manifold with $H^{(2,0)}(M)\not=0$ the perturbed Lagrangian $\Delta
L$ can be rewritten in terms of $\omega$ being a non-zero
holomorphic form in $H^{(2,0)}(M)$ such that it vanishes on $C$.
Consequently, the mass term vanishes precisely in the zeros of
$\omega$ over the divisor (global cosmic strings)\footnote{This
cosmic string indeed captures chiral fermion  zero modes of the
field $\psi$ which propagates on the canonical divisor $C$.}.

Assume that ${C}=\bigcup_y {C}_y$, where ${C}_y$ is a Riemann
surface for each $y$ such that $\omega$ has at most simple zeroes on
$C_y$. One would estimate the contribution of the divisor (cosmic
string) to the Donaldson-Witten invariants by considering the
intersections ${\Sigma} \cap C_y \not= \emptyset$. In the
intersection (which is assumed to be transverse) points $P$ one can
insert operators $V_y(P)$. Thus, if $\#(\Sigma \cap C_y)$ is the
intersection number of $\Sigma$ and $C_y$ is given by
\begin{equation}
\# (\Sigma \cap C_y ) = \int_{M} \theta_\Sigma \wedge \theta_{C_y},
\label{intCwS}
\end{equation}
where $\theta_\Sigma$ is the Poincar\'e dual of $\Sigma$ and
$\theta_{C_y}$ is the Poincar\'e dual of $C_y$. Then the operators
${I}(\Sigma)$ must be replaced by $\sum_y \#({\Sigma} \cap C_y)V_y$.
Here $V_y=V_y(P)$ is a local operator inserted in the intersection
points $P$'s between $\Sigma$ and $C_y$.

For the theory on the worldsheet (cosmic string) $C$ it is assumed
that it has a mass gap and a chiral symmetry breaking with vacuum
degeneracy determined by $\IZ_2$. Fermionic zero modes on the
divisor lead also to a non-vanishing anomaly inside the theory on
$C_y$ which should cancel by other trapped fields along the string.
Thus these chiral fermions contributes to the path integral measure
by a factor $t_y=(-1)^{d \varepsilon_y}$ where $\varepsilon_y=0, 1$
and $d$ is the dimension of the gauge group. Gathering all together
in Ref. \cite{MINIMAL} it was found that the Donaldson-Witten
invariant is of the form
$$
\bigg{\langle} {\exp} \bigg( \sum_a \alpha_a {I}(\Sigma) + {\lambda}
{\cal O}\bigg) \bigg \rangle
$$
$$ = 2^{{1\over 4}(7\chi + 11 \sigma)}{\exp} \bigg( {1\over 2}
\sum_{a,b} {\alpha}_a {\alpha}_b \#({\Sigma}_a \cap {\Sigma}_b) + 2 {\lambda}  \bigg)  \cdot \prod_y
\bigg(e^{{\phi}_y} + {t}_y
e^{-{\phi}_y}\bigg)
$$
\begin{equation}
+ i^{\Delta}2^{1+{1\over 4}(7\chi + 11 \sigma)}{\exp} \bigg(
-{1\over 2} \sum_{a,b} {\alpha}_a {\alpha}_b \#({\Sigma}_a \cap
{\Sigma}_b) - 2 {\lambda} \bigg) \cdot \prod_y \bigg(e^{-{\phi}_y} +
{t}_y e^{{\phi}_y}\bigg), \label{dwinvariantsforK}
\end{equation}
where
\begin{equation}
{\phi}_y = \sum_a {\alpha}_a \#({\Sigma}_a\cap C_y)
\end{equation}
and $\Delta = {1 \over 2} d({\cal M}_D)$.

The invariants (\ref{dwinvariantsforK}) can be further generalized in
the case the divisor components $C_y$ have singularities. Also the
consideration of higher dimensional gauge groups leads to
interesting generalizations. Both extensions were discussed in Ref.
\cite{MINIMAL}.

In summary, in the process of the obtention of
(\ref{dwinvariantsforK}) there were made a series of physical
considerations. It was assumed cluster decomposition with a set of
vacuum states, mass gap, chiral symmetry breaking, the smooth
breaking of supersymmetry by the introduction of a mass term in the
matter multiplet. All these assumptions are reasonable except the
mass gap. It is well known that supersymmetric ${\cal N}=2$
Yang-Mills theory don't have a mass gap. However this assumption
make sense as one adds terms in the Lagrangian of the original
${\cal N}=2$ theory which leaves only one unbroken supersymmetry.
The theory is ${\cal N}=1$, the mass gap is allowed and it gives
precisely the necessary ingredient to interpret
(\ref{dwinvariantsforK}) as the Donaldson-Witten invariants of
K\"ahler manifolds. This is the subject of the following subsection.

\subsection{Asymptotic Observables in K\"ahler Manifolds}

Before we proceed with the case of K\"ahler manifolds we make some
considerations of general character about the asymptotic
intersection of two asymptotic cycles. When a complex structure is
defined in $M$, every form and vector field can be in general
decomposed in a holomorphic, mixed and anti-holomorphic parts. Let
$(z^n,{z}^{\overline{n}})$ be complex coordinates on $M$ and
$\{{dz^n, d{z}^{\overline{n}}}\}$ a basis for the cotangent space
$T_x^\ast M$ and $\{{\partial_n, \partial_{\overline{n}}}\}$ a basis
for the tangent space $T_x M$ at the point $x$. Then we can
decompose our observable as \cite{Park:1993fy}:
\begin{equation}
Z = Z^{(0,2)} + Z^{(1,1)} + Z^{(2,0)},
\end{equation}
where in complex coordinates it looks like $Z^{(0,2)}= Z_{mn} dz^{m}
\wedge dz^{n}$, $Z^{(1,1)} = Z_{m \overline{n}} dz^{n} \wedge
dz^{\overline{n}}$ and $Z^{(2,0)} = Z_{\overline{m} \overline{n}}
dz^{\overline{m}} \wedge dz^{\overline{n}}$. In general every
element corresponds to the decomposition of $\Omega^p (M)
=\oplus_{p=r+s} \Omega^{(r,s)} (M)$, where $r$ and $s$ stands for
the degrees of the corresponding holomorphic and anti-holomorphic
components. One has a direct sum decomposition of $p$-vector fields
$\mathcal{H}^p (M) = \oplus_{p=r+s} \mathcal{H}^{(r,s)} (M)$. Thus
for $p=2$:
\begin{equation}
{\bf Y} = {\bf Y}^{(0,2)} + {\bf Y}^{(1,1)} + {\bf Y}^{(2,0)},
\end{equation}
where ${\bf Y}^{(0,2)} = Y^{mn} \partial_{{m}} \wedge
\partial_{{n}}$, ${\bf Y}^{(1,1)} = Y^{m \overline{n}} \partial_{n}
\wedge \partial_{\overline{n}}$ and ${\bf Y}^{(2,0)}
=Y^{\overline{m} \overline{n}}  \partial_{\overline{m}} \wedge
\partial_{\overline{n}}$. Now a $2$-vector field can be constructed
from vector fields as ${\bf Y} = Y_{1} \wedge Y_{2}$, where each
$Y_{i}$ (for $i=1,2$) is a vector field. Each $Y_i$ can be
decomposed as the sum of a holomorphic and a anti-holomorphic part
as follows, $Y_{i}=Y_{i}^{m} \partial_{m} + Y_{i}^{\overline{m}}
\partial_{\overline{m}}$ then the $2$-vector field ${\bf Y}$ takes
the form
\begin{equation}
{\bf Y} = Y_{1}^{m} Y_{2}^{n} \partial_{m} \wedge \partial_{n} +( Y_{1}^{m} Y_{2}^{\overline{n}}- Y_{1}^{\overline{n}} Y_{2}^{m}) \partial_{m} \wedge \partial_{\overline{n}}  + Y_{1}^{\overline{m}} Y_{2}^{\overline{n}} \partial_{\overline{m}} \wedge \partial_{\overline{n}}.
\end{equation}
Thus we can calculate the contraction of the $2$-vector field ${\bf
Y}$ and the observable $Z$. By the orthogonality relations between
the basis, the asymptotic observable takes the following form
$$
\widetilde{I}_{{\bf Y}}(\mu) = \int_{M} \left[i_{{\bf
Y}^{(0,2)}}(Z^{(0,2)})   + i_{{\bf Y}^{(1,1)}} (Z^{(1,1)})   +
i_{{\bf Y}^{(2,0)}} (Z^{(2,0)})  \right] \mu
$$
\begin{equation}
\label{kobsoy} = \int_{M} \left[ Z_{mn} \cdot Y^{mn} + Z_{m
\overline{n}} \cdot Y^{m \overline{n}} + Z_{\overline{m}
\overline{n}} \cdot Y^{\overline{m}\overline{n}} \right] \mu.
\end{equation}

If the 2-vector fields are coming from the product of vector fields
we have
\begin{equation}
\label{kobs} \widetilde{I}_{{\bf Y}}(\mu)  = \int_{M} \left[ Z_{mn}
\cdot Y_1^{m} Y_{2}^{n}  + Z_{m \overline{n}} \cdot ( Y_{1}^{m}
Y_{2}^{\overline{n}}- Y_{1}^{\overline{n}} Y_{2}^{m})   +
Z_{\overline{m} \overline{n}} \cdot Y_{1}^{\overline{m}}
Y_{2}^{\overline{n}}  \right] \mu.
\end{equation}
 Moreover we can decompose the asymptotic observables into three different parts,
 one associated to a completely holomorphic part, one to a completely anti-holomorphic
 part and one to the mixed component.

For physical reasons \cite{MINIMAL} there are only three type of
relevant observables. These are given by:

\begin{itemize}
\item{} The usual observable $\widetilde{I}(\Sigma)$ that contributes
to the asymptotic intersection number (\ref{internumb}) is given by:
$\int_M i_{{\bf Y}^{(2,0)}} (Z^{(2,0)})  \mu + {\rm h.c.}$, with
$Z^{(2,0)}$ given in (\ref{observa}).

\item{} The observable $I(\omega)= \int_\Sigma \omega$ arises
from a non-vanishing mass term which breaks supersymmetry (from
${\cal N}=2$ to ${\cal N}=1$). The asymptotic version is written as
$\int_M i_{{\bf Y}^{(2,0)}} (\omega^{(2,0)}) \mu + {\rm h.c.}$ and
it contributes to the asymptotic intersection number (\ref{intCwS}).

\item{} Near intersection points of $\Sigma$
and $C_y$, where it is assumed to be inserted an operator $V(P)$
gives a term of the form $I(\theta)= \int_M \theta \wedge Z$. The
natural form that couples to the asymptotic canonical divisor
$\widetilde{C}_y$ is given by the first Chern class
$J=c_1(\widetilde{C}_y)$ and it is a 2-form of type $(1,1)$. It can
be associated to a vector field ${\bf X}^{(1,1)}$. Thus we have that
the observable is given by  $\int_M  i_{{\bf X}^{(1,1)}} (J) \mu$.

\end{itemize}

\subsection{Invariants for K\"ahler Manifolds }

Let $\widetilde{C}$ be  the disjoint union of a finite number of
$\widetilde{C}_y$, where $\widetilde{C}_y$ is an asymptotic Riemann
surface for each $y$ with simple zeroes. This asymptotic cycle can
be defined in terms a free divergence 2-vector field ${\bf X}$ and
it is given by:
\begin{equation}
\widetilde{C}_y({\bf X}) = \int_M i_{{\bf X}} (J)  \mu,
\end{equation}
where $J=c_1(C_y)$ and ${\bf X}\in {\cal H}^{(1,1)}$.

One can estimate the contribution of the divisor (cosmic string) to
the Donaldson-Witten invariants of $M$ with flows. The contribution
of the intersections $\widetilde{\Sigma}_{\bf Y} \cap
\widetilde{C}_y \not= \emptyset$ can be computed as follows. The
operators $\widetilde{I}_{{\bf Y}}(\mu)(x)$, inserted on the
canonical divisor contribute precisely to the intersections of the
cycles $\widetilde{\Sigma}$ with the canonical divisor
$\widetilde{C}_y$ multiplied by a  local operator $V(P)$. Thus it
should make the replacement:
\begin{equation}
\widetilde{I}_{{\bf Y}} (\mu)(x) \to \sum_y
\#(\widetilde{\Sigma}_{\bf Y} \cap \widetilde{C}_y)V_y + {\rm terms
\ involving \ intersections  \ of \ \Sigma's} .
\end{equation}
The expectation values in the vacua of operators $V(P)$ are fixed by
normalization. Thus the main contribution comes from
$\#(\widetilde{\Sigma}_{\bf Y} \cap \widetilde{C}_y)$ which is given
by the intersection number of two flows, one of them associated to
the $\widetilde{\Sigma}$ and the other one to the asymptotic
canonical divisor $\widetilde{C}_y$
\begin{equation}
\# (\widetilde{\Sigma}_{\bf Y} \cap \widetilde{C}_y ) = \int_{M_1
\times M_2} (Z\wedge \Theta_{\bf Y})(x_1) \wedge (J \wedge
\Theta_{\bf X})(x_2) \cdot \delta(x_1-x_2), \label{intCwStwo}
\end{equation}
where $\Theta_{\bf Y} = i_{{\bf Y}} (\mu)$ and  $\Theta_{\bf
X}=i_{{\bf X}}(\mu)$ are the Poincar\'e duals of
$\widetilde{\Sigma}_{\bf Y}$ and $\widetilde{C}_y$ respectively.
Here the measure $\mu$ is also invariant under the flow ${\bf X}$
generating the canonical divisor.

Gathering all together the previous considerations we have a form
for the Donaldson-Witten invariants for K\"ahler manifolds with
canonical divisor:
$$
\bigg{\langle} {\exp} \bigg( \sum_a \alpha_a \widetilde{I}_{{\bf Y}}
(\mu) + {\lambda} {\cal O}\bigg) \bigg \rangle
$$
$$ = 2^{{1\over 4}(7\chi + 11 \sigma)}{\exp} \bigg( {1\over 2}
\sum_{a,b} {\alpha}_a {\alpha}_b \#(\widetilde{\Sigma}_{{\bf Y}_a} \cap
\widetilde{\Sigma}_{{\bf Y}_b}) + 2 {\lambda}  \bigg)  \cdot \prod_y
\bigg(e^{\widetilde{\phi}_y} + {t}_y
e^{-\widetilde{\phi}_y}\bigg)
$$
\begin{equation}
+ i^{\Delta}2^{1+{1\over 4}(7\chi + 11 \sigma)}{\exp} \bigg(
-{1\over 2} \sum_{a,b} {\alpha}_a {\alpha}_b
\#(\widetilde{\Sigma}_{{\bf Y}_a} \cap \widetilde{\Sigma}_{{\bf Y}_b}) - 2 {\lambda}
\bigg) \cdot \prod_y \bigg(e^{-\widetilde{\phi}_y} +
{t}_y e^{\widetilde{\phi}_y}\bigg).
\end{equation}
where $\widetilde{\phi}_y$ is given by Eq. (\ref{intCwStwo}) and
$\Delta = {1 \over 2} d({\cal M}_D)$.

Finally, it is worth mentioning that one can generalize these
expressions for more general K\"ahler manifolds with canonical
divisors that don't have simple zeros. Further generalizations to
non-simply connected manifolds, with $\pi_1(M) \not= 0$ and for
higher dimensional gauge groups the reader can see, for instance
\cite{Argyres:1994xh}.

\subsection{Examples}
\begin{enumerate}
\item {\bf Flows on hyper-K\"ahler manifolds with
{\boldmath $H^{(2,0)}(M) \not= 0$}.} We start with
$r$ divergence-free $2$-vector fields $Y_a$ with $a=1,\ldots, r$ on
a 4-torus ${\bf T}^4$. Then the asymptotic invariants associated to
every $2$-vector field can be computed through the correlation
functions of a product of operators $\widetilde{I}({\bf Y}_a) =
\int_M i_{{\bf Y}_a}(Z^{(2,0)})  \mu_a + {\rm h.c} = \int_M i_{{\bf
Y}_a}(\lambda \psi + \overline{\omega}\overline{B}B) \mu_a + {\rm
h.c.}$. Thus to this configurations of flows we get the following
invariant by using the Hodge structure of the torus: $h^{1,0}=2$,
$h^{2,0}=1$. The invariant is given by
\begin{align}
\label{kahleronetwo}
\bigg\langle \exp \bigg( \sum_a \alpha_a \tilde I (Y_a) + \lambda \mathcal{O} \bigg) \bigg\rangle & = \nonumber  \exp \bigg( \frac{1}{2} \sum_{a,b} \alpha_a \alpha_b \#(\widetilde{\Sigma}_{{\bf Y}_a} \cap \widetilde{\Sigma}_{{\bf Y}_b}) + 2 \lambda \bigg) \notag \\
&+ \exp \bigg( -\frac{1}{2} \sum_{a,b} \alpha_a \alpha_b \#(\widetilde{\Sigma}_{{\bf Y}_a} \cap \widetilde{\Sigma}_{{\bf Y}_b}) - 2 \lambda \bigg)
\end{align}
where $\#(\widetilde{\Sigma}_{{\bf Y}_a} \cap
\widetilde{\Sigma}_{{\bf Y}_b})$ is the asymptotic intersection
number. Thus we find that $\#(\widetilde{\Sigma}_{{\bf Y}_a} \cap
\widetilde{\Sigma}_{{\bf Y}_b})$ is given by Eq.
(\ref{intersecdos}). In the specific case of ${\bf T}^4$ we have 2
cycles of dimension two, one holomorphic and the other
anti-holomorphic. The vector fields ${\bf Y}_a$ are wrapped on these
homology cycles. There is one-dimensional homology cycles and one
can introduce vector fields whose orbits coincide with these cycles.
One can construct asymptotic invariants associated with them and
compute their contribution to the correlation functions. But we are
not interested in this addition in the present paper. However it is
interesting to remark that 2-vector fields can be constructed from
the wedge product of two of these vector fields and we have
constructed observables by using (\ref{kobs}). Thus we find that our
invariant can be computed by using Eq. (\ref{internumb}).

For the case of K3, where $h^{1,0}=0$, $h^{2,0}=1$, we don't have
nonsingular vector fields (since the Euler characteristic is 24).
However we have only intrinsic 2-vector fields. Thus there will be
not 2-vector fields constructed from 1-vector fields as in the
previous example. In this case the invariant is given by

\begin{align}
\label{kahler} \bigg\langle \exp \bigg( \sum_a \alpha_a \tilde I
(Y_a) + \lambda \mathcal{O} \bigg) \bigg\rangle & =
 \nonumber  \notag
C \bigg[\exp \bigg( \frac{1}{2} \sum_{a,b} \alpha_a \alpha_b
\#(\widetilde{\Sigma}_{{\bf Y}_a} \cap \widetilde{\Sigma}_{{\bf
Y}_b}) +
 2 \lambda \bigg)\\
& - \exp \bigg( -\frac{1}{2} \sum_{a,b}
\alpha_a \alpha_b \#(\widetilde{\Sigma}_{{\bf Y}_a} \cap
\widetilde{\Sigma}_{{\bf Y}_b}) - 2 \lambda \bigg)\bigg]
\end{align}
where $C=1/4$. The asymptotic intersection number is also given only
by Eq. (\ref{intersecdos}) and the sum involves also normal and
asymptotic self-intersection numbers.

\item {\bf Hilbert Modular surfaces} \cite{VderGeer}. Let $\H\subset \C$ denote the upper half-plane with the Poincar\'e metric.
Let $K:=\Q(\sqrt{2})$ be the totally real quadratic number field obtained by adding $\sqrt2$ to $\Q$.
The ring of integers $\Z(\sqrt2)$ is the ring of real numbers of the form $m+n\sqrt2$, $\,m,n \in\Z$.
Let $\sigma$ be the nontrivial Galois automorphism
of $K$ given explicitly by $\sigma(a+b\sqrt{2})=a-b\sqrt{2}$, $\,\,a, b\in\Q$.
Let us consider the group $\Gamma:=PSL(2,\Z(\sqrt2))$. Let $\bar\sigma:\Gamma\to \Gamma$ be the
induced automorphism on $\Gamma$.

$\Gamma$ acts properly and discontinuously on
$\H\times\H$ as follows: $\gamma(z,w)=(\gamma(z),\sigma(\gamma)(w))$.
The quotient is an orbifold  of dimension four which is not compact but it has finite volume
(with respect to the induced metric coming from the product metric on $\H \times\H$).
The ends (cusps) of Hilbert modular surfaces of real quadratic number fields are manifolds which
are of form $M^3\times [0,\infty)$ where $M^3$ is a compact solvable 3-manifold which fibers over
the circle with
fibre a torus $\mathbb T^2$. The number of such ends is equal to the class number of the field
\cite{VderGeer}.
For $\Q(\sqrt2)$ there is only one cusp and $M^3$ is the mapping torus of the automorphism of the 2-torus induced by the matrix
$A=\begin{pmatrix}
2&1\\
1&1
\end{pmatrix}$. Thus $M^3$ fibers over the circle. In fact, the sugbroup $\Lambda\subset\Gamma$
consisting of affine transformations of the form
$z\mapsto(1+\sqrt2)^{r}z+m+n\sqrt2$, $\,r,m,n \in\Z$ is the
semidirect product $\Z\ltimes_A(\Z\times\Z)$ which is the solvable
fundamental group of $M^3$. The universal cover of $M^3$ is a
solvable simply connected 3-dimensional Lie group whose Lie algebra
is generated by three left-invariant vector fields ${\bf X}$, ${\bf
Y}$ and ${\bf Z}$ whose Lie brackets satisfy $[{\bf X},{\bf
Y}]=a{\bf Y}$, $[{\bf X},{\bf Z}]=-a{\bf Z}$ and $[{\bf Y},{\bf
Z}]=0$, for some constant $a>0$. The commuting vector fields ${\bf
Y}$ and ${\bf Z}$ descend to vector fields in $M^3$ which are
tangent to the torus fibers of the fibration of $M^3$ over the
circle. The flow generated by ${\bf X}$ is an Anosov flow. The
vector fields ${\bf Y}$ and ${\bf Z}$ generate two flows tangent to
1-dimensional foliations $L_1$ and $L_2$. These flows are homologous
to zero and the Arnold's self-linking number of both is zero.

By a theorem of Selberg  \cite{Selberg} $\Gamma$ contains a finite
index subgroup $\tilde\Gamma$ which acts freely on $\H\times\H$. The
quotient manifold $M^4(\tilde\Gamma):=\tilde\Gamma/\H\times\H$ is a
non compact manifold of finite volume with a finite number of cusps
depending upon the Selberg subgroup. The action of
$PSL(2,\Z(\sqrt2))$ preserves the natural foliations of $\H \times
\H$ whose leaves are, respectively, of the form $\H\times\{w\}$ and
$\{z\}\times\H$. These foliations descend to $M^4(\tilde\Gamma)$ to
a pair of 2-dimensional foliations ${\mathcal F}_{horizontal}$ and
${\mathcal F}_{vertical}$ which are mutually transverse and each has
dense leaves. Furthermore, since the action of $\Gamma$ is by
isometries, the foliations ${\mathcal F}_{horizontal}$ and
${\mathcal F}_{vertical}$ are transversally Riemannian and thus both
have natural transverse measures.

Now we can do two things to obtain examples of 4-manifolds with (possibly singular) foliations.
\begin{itemize}
\item We can compactify $M^4(\tilde\Gamma)$ \`a la Hirzebruch \cite{Hirzebruch}
by adding one point at infinity for each cusp. The resulting space is an algebraic
surface with singularities
at the cusps and the link of each singularity is the corresponding solvmanifold.
After desingularizing one obtains a smooth algebraic surface which is therefore a K\"ahler surface.
The foliations ${\mathcal F}_{horizontal}$
and ${\mathcal F}_{vertical}$ lift to the desingularized manifold to foliations with
singularities at the cusps.
There are important relations of these constructions with $K3$ surfaces \cite{FM}.
\item One can ``cut'' the manifold at each cusp, to obtain a compact manifold with boundary and each
component of the boundary is a solvmanifold described before. In other words, we remove
a conic open neighborhood of each cusp whose boundary is the corresponding solvmanifold
at the cusp.  Now we can take the double to obtain a compact
closed manifold with a pair of transversally Riemannian foliations with dense leaves
(since in the double the foliations can be glued differentiably).  Both foliations
meet transversally the solvmanifolds and determine two flows in them.
For the case $K=\Q(\sqrt2)$ we obtain a compact 4-manifold with two transversally
Riemannian foliations which meet the solvmanifold in the foliations $L_1$ and $L_2$
above. Therefore: {\it each of the foliations ${\mathcal F}_{horizontal}$
and ${\mathcal F}_{vertical}$ has self-intersection zero} (since $L_1$ and $L_2$
have self intersection zero).

\end{itemize}

Of course one can construct examples as above using any totally real quadratic field and the group
$PSL(2,{\mathfrak{O}}_K)$, where  ${\mathfrak{O}}_K$ is the ring of integers of $K$.

\item {\bf Elliptic K3 surfaces end elliptic surfaces}. Let $S$ be an elliptic surface with
Kodaira fibration $\pi:S\to \Sigma_g$, where $\Sigma_g$ is an
algebraic curve of genus $g$. The fibres are elliptic curves except
for a finite number of singular fibres which are rational curves.
The fibration provides us with a singular foliation as mentioned in
the remark above. There is a canonical choice for a transverse
measure $\mu$ which is obtained from the Poincar\'e metric via the
uniformization theorem applied to $\Sigma_g$: if $\tau$ is a 2-disk
which is transversal to the regular part of the foliation its
measure is the hyperbolic area of $\pi(\tau)$. Then we can apply our
results to the triple $(S, \mathcal{F}, \mu)$. One modification of
elliptic surfaces can be obtained by the so-called {\it logarithmic
transformation}. Using logarithmic transformation one can change the
Kodaira dimension and turn an algebraic surface into a non algebraic
surface.

\noindent Particular cases of elliptic surfaces
are the $K$3 surfaces, Enriques surfaces and the Dolgachev surfaces.
We recall that Dolgachev $X_p$ surfaces depend on an integer $p$
were used by Donaldson to obtain the first examples $X_2$ and $X_3$ of manifolds which are
homeomorphic but not diffeomorphic  \cite{FM}, \cite{DK}.
From the above two questions arise:
\begin{itemize}

\item How do our invariants change after performing a logarithmic transformation on an elliptic surface?

\item Can we detect exotic differentiable structures  by our invariants?

\end{itemize}

\item {\bf Symplectic 4-manifolds and Lefschetz fibrations and pencils}. By a result of Donaldson
\cite{pencil} every symplectic 4-manifold admits a Lefschetz fibration and
these fibrations are an essential tool for the study of symplectic 4-manifolds.
As the previous example, one has a triple $(M^4, \mathcal{F}, \mu)$, where $\mathcal{F}$ is the
(possibly singular) foliation determined by the Lefschetz fibration and $\mu$ is a transverse
measure coming from a choice of an area form from a Riemannian metric on the base surface.
The question is how to compute our invariants and how can they be used to study symplectic manifolds.
\end{enumerate}

\section{Survey on Seiberg-Witten Invariants}

Another example of the theories which can be constructed through the
Mathai-Quillen formalism is the cohomological field theory
describing Seiberg-Witten monopoles \cite{MONO}. The geometric data
consists of the square root of the (determinant) line bundle
$L^{1/2}$ over a four-manifold $M$ with an abelian gauge connection
$A$ with curvature $F_A=dA$. We have also the tensor product $S^\pm
\otimes L^{1/2}$ of $L^{1/2}$ with the spin bundle $S^\pm$, which
exist whenever $M$ is a spin manifold i.e. $w_2(M)=0$ (for more
details on the spin structure see, \cite{SPIN,SPINSALAMON}). This
tensor product is even well defined if $M$ is not a spin manifold.
In addition we have a section $\psi_\alpha \in \Gamma(S^+ \otimes
L^{1/2})$. The Seiberg-Witten equations are

\begin{equation}
F^+_{\alpha \beta} = - {i \over 2} \overline{\psi}_{(\alpha}\psi_{\beta)}, \ \ \ \ \ \ D_{\alpha \dot{\alpha}}\psi^\alpha =0,
\end{equation}
where $F^+$ is the self-dual part of the curvature $F_A$. Here
$\alpha,\beta$ are spinorial indices instead of vector ones $\mu$
and they are related by $A_\mu = \sigma_\mu^{\alpha
\dot{\alpha}}A_{\alpha \dot{\alpha}}$.

The moduli space ${\cal M}_{SW}$ of solutions to the Seiberg-Witten
equations will be denoted as ${\cal M}_{SW} \subset {\cal A} \times
\Gamma(S^+ \otimes L^{1/2})/{\cal G}$, where ${\cal A}$ is the space
of abelian connections on $L^{1/2}$ and ${\cal G}$ is the gauge
group of the $U(1)$-bundle, i.e. ${\cal G}= {\rm Map}(M,U(1))$. This
moduli problem can be described in terms of the Mathai-Quillen
construction. In this case the vector bundle is also trivial ${\cal
V}= {\cal M} \times {\cal F}$, where ${\cal F}$ is the fibre. For
the monopole case ${\cal F}= \Lambda^{2,+}(M) \otimes \Gamma(S^-
\otimes L^{1/2})$. The section $s$ is given by

\begin{equation}
s(A,\psi)= \bigg({1 \over \sqrt{2}}\big(F^+_{\alpha \beta} + {i \over 2} \overline{\psi}_{(\alpha}\psi_{\beta)}\big), D_{\alpha \dot{\alpha}} \psi^\alpha  \bigg),
\end{equation}
where $D_{\alpha \dot{\alpha}} \psi_\beta = \sigma^\mu_{\alpha
\dot{\alpha}} (\partial_\mu + i A_\mu)\psi_\beta$. The zero section
determines precisely the Seiberg-Witten equations.

The dimension $d({\cal M}_{SW})$ of the moduli space ${\cal M}_{SW}$
can be obtained from an index theorem
\begin{equation}
 d({\cal M}_{SW}) = \lambda^2 - {2 \chi + 3 \sigma \over 4},
\end{equation}
where $\lambda = {1 \over 2} c_1(L)$ (being $c_1(L)$ the first Chern
class), $\chi$ and $\sigma$ are the Euler characteristic and the
signature of $M$ respectively. The Mathai-Quillen construction
\cite{MQ} provides with a set of fields  $A_\mu$, $\psi_\mu$,
$\phi$, $\chi_{\mu \nu}$, $H_{\mu \nu}$, $\eta$, $\psi_\alpha$,
$\mu_\alpha$, $v_{\dot{\alpha}}$ and $h_{\dot{\alpha}}$ of different
ghost number. This set of fields will be denoted for short as
${\cal X}$. The Lagrangian can be read off from the exponential of
the Thom class and is given by \cite{book1,Labastida:1995bs}
$$
L_{SW} = \int_M e \bigg(g^{\mu \nu} D_\mu \overline{\psi}^\alpha D_\nu \psi_\alpha + {1 \over 4} R \overline{\psi}^\alpha \psi_\alpha + {1 \over 2} F^{+ \alpha \beta} F^+_{\alpha \beta} -
{1 \over 8} \overline{\psi}^{(\alpha} \psi^{\beta)} \overline{\psi}_{(\alpha} \psi_{\beta)}
\bigg)
$$
$$
+ i \int_M \bigg( \lambda \wedge * d^*d \phi - {1 \over \sqrt{2}} \chi \wedge * \rho^+ d \psi \bigg)
+ \int_M \bigg[ i \phi \lambda \overline{\psi}^{\alpha} \psi_{\alpha} + {1 \over 2 \sqrt{2}} \chi^{\alpha \beta} (\overline{\psi}_{(\alpha} \mu_{\beta)} + \overline{\mu}_{(\alpha} \psi_{\beta)})
$$
\begin{equation}
- {i \over 2}(v^{\dot{\alpha}} D_{\alpha \dot{\alpha}} \mu^\alpha - \mu^{\alpha} D_{\alpha \dot{\alpha}} v^{\dot{\alpha}})
- {i \over2} [\overline{\psi}^\alpha \psi_{\alpha \dot{\alpha}} v^{\dot{\alpha}}] +
{1\over 2} \eta (\overline{\mu}^\alpha \psi_\alpha) - \overline{\psi}^\alpha \mu_\alpha)
+ {i \over 4} \phi \overline{v}^{\dot{\alpha}} v_{\dot{\alpha}} - \lambda \overline{\mu}^\alpha \mu_\alpha \bigg].
\end{equation}

The observables are products of BRST invariant operators which are
cohomologically non-trivial
\begin{equation}
\label{ebrst}
d \Theta^n_p = \{ {\cal Q},\Theta_{p+1}^n \}.
\end{equation}
The ${\cal Q}$-invariant operators are \cite{Labastida:1995bs}
$$
{\cal O}^{\gamma_0}_n = \Theta_0^n(x)
$$
$$
{\cal O}^{\gamma_1}_n = \int_{\gamma_1} \Theta_1^n, \   \ \ \ \ \ \ \
{\cal O}_n^{\gamma_2} = \int_{\gamma_2} \Theta_2^n
$$
\begin{equation}
\label{obsw}
{\cal O}_n^{\gamma_3} = \int_{\gamma_3} \Theta_3^n, \   \ \ \ \ \ \ \
{\cal O}_n^{\gamma_4} = \int_{M} \Theta_4^n,
\end{equation}
where
$$
\Theta_0^n = \left(
\begin{array}{cc}
  n \\
  0 \\
\end{array} \right) \phi^n, \ \ \ \ \ \ \ \
\Theta_1^n = \left(
\begin{array}{cc}
  n\\
  1 \\
\end{array} \right) \phi^{n-1} \psi,
$$
$$
\Theta_2^n = \left(
\begin{array}{cc}
  n \\
  2 \\
\end{array} \right) \phi^{n-2} \psi \wedge \psi + \left(
\begin{array}{cc}
  n\\
  1 \\
\end{array} \right) \phi^{n-1}  \psi \wedge F,
$$
$$
\Theta_3^n =  \left(
\begin{array}{cc}
  n\\
  3 \\
\end{array} \right)\phi^{n-3} \psi \wedge \psi \wedge \psi + 2\left(
\begin{array}{cc}
  n\\
  2 \\
\end{array} \right) \phi^{n-2}  \psi \wedge F,
$$
\begin{equation}
\label{formsw}
\Theta_4^n = \left(
\begin{array}{cc}
  n\\
  4 \\
\end{array} \right) \phi^{n-4} \psi \wedge \psi \wedge \psi \wedge \psi + 3 \left(
\begin{array}{cc}
  n\\
  3 \\
\end{array} \right)\phi^{n-3}  \psi \wedge \psi \wedge F + \left(
\begin{array}{cc}
  n\\
  2 \\
\end{array} \right) \phi^{n-2} F \wedge F.
\end{equation}

Here $\Theta_0^n(x)$ is constructed with a gauge and ${\cal Q}$
invariant field $\phi$. All other observables are descendants
obtained from it \cite{Labastida:1995bs}. As in the Donaldson-Witten
case the construction establishes an isomorphism between the BRST
cohomology $H^*_{BRST}({\cal Q})$ and the de Rham cohomology
$H^*_{dR}(M)$. To be more precise the analogue to the Donaldson map
is: $\delta_{SW}: H_p(M) \to H^{2-p}({\cal M}_{SW})$ given in terms
of the first Chern class of ${\cal V}$. The observables (\ref{obsw})
are BRST invariant (BRST closed) and the BRST commutator only
depends of the homology class. This can be shown by following a
similar procedure as we did in the Donaldson case (see Eqs.
(\ref{ecuaciondiez}) and (\ref{ecuaciononce})).

The correlation functions of $r$ operators are written as
\begin{eqnarray}
 \big\langle \mathcal{O}_n^{\gamma_{p_1}} \cdots
\mathcal{O}_n^{\gamma_{p_r}} \big\rangle & = &  \bigg\langle  \prod_{j=1}^r \int_{{\gamma}_{p_j}}
\Theta^n_{p_j} \bigg\rangle \nonumber \\
& = & \int  \mathcal{D} \mathcal{X}
\exp(-L_{SW}/e^2) \prod_{j=1}^r \int_{\gamma_{p_j}}
\Theta^n_{p_j} .
\label{eq:inv13}
\end{eqnarray}
These are the Seiberg-Witten invariants in the path integral
representation \cite{book1,Labastida:1995bs}. They are topological
invariants and also invariants of the smooth structure of $M$. After
integration over the non-zero modes one has:
\begin{eqnarray}
 \big\langle \mathcal{O}_1^{\gamma_{p_1}} \cdots
\mathcal{O}_r^{\gamma_{p_r}} \big\rangle & = &  \bigg\langle
\prod_{j=1}^r \int_{{\gamma}_{p_j}}
\Theta^n_{p_j} \bigg\rangle \nonumber \\
& = & \int_{{\cal M}_{SW}} \nu_{p_1} \wedge \cdots \wedge \nu_{p_r},
\label{eq:inevitabletwo}
\end{eqnarray}
where $\nu_{p_j}= \delta_{SW}(\gamma_{p_j})$. The possible values of
$p_j$ are $0,1,2$.  Thus, for $p_j =1,2$, we can rewrite the
previous equation as
\begin{equation}
 \big\langle \mathcal{O}_1^{\gamma_{1}} \cdots
\mathcal{O}_r^{\gamma_{1}} \cdot \mathcal{O}_1^{\gamma_{2}} \cdots
\mathcal{O}_{d/2}^{\gamma_{2}}\big\rangle  =  \int_{{\cal M}_{SW}}
\nu_{{1}_1} \wedge \cdots \wedge \nu_{{1}_r} \wedge
\phi_\Sigma^{d/2}, \label{swinv}
\end{equation}
where $\phi_\Sigma$ are 2-forms on the SW moduli space. For simply
connected manifolds $\pi_1(M)=0$ the relevant cycles are of
dimension $p=2$. The Seiberg-Witten invariants can be also written
in terms of differential forms in the moduli space ${\cal M}_{SW}$
in the form \cite{book1,Labastida:1995bs}
\begin{equation}
 \big\langle \mathcal{O}_1^{\gamma_{2}} \cdots
\mathcal{O}_{d/2}^{\gamma_{2}} \big\rangle  = \int_{{\cal M}_{SW}}\phi_\Sigma^{d/2} .
\label{eq:inevitableon}
\end{equation}

\section{Seiberg-Witten Invariants for Flows}

In order to incorporate flows in the Seiberg-Witten theory we define
\begin{equation}
\label{eq:asin} \widetilde{\cal O}_{{\bf Y}_p}(\mu)  = \int_M
i_{{\bf Y}_p}(\Theta^n_p)\mu_p(x) ,
\end{equation}
where $\mu_p$ is the invariant volume form. We can interpret the
integral as the averaged asymptotic cycles on $M$ by the Schwartzman
theorem \cite{SH}, $i_{{\bf Y}_p} (\mu)$ is a closed $(4-p)$-form,
from which we will obtain a asymptotic $p$-cycle
$\widetilde{\gamma}_p$ by Poincar\'e duality (an element of the
$H_p(M,\mathbb{R})$).

Let ${\bf Y}_p$ be  $p$-vector fields with $p=0,1,2$, then the
expression $(\ref{eq:asin})$ defines the asymptotic observables as
\begin{eqnarray}
 \widetilde{\mathcal{O}}^n_{ {\bf Y}_1}(\mu_1) =  \int_M
 i_{{\bf Y}_1} (\Theta^n_1) \,\mu_1,\ \ \ \ \ \
 \widetilde{\mathcal{O}}^n_{{\bf Y}_2}(\mu_2) = \int_M i_{{\bf Y}_2} (\Theta^n_2) \, \mu_2,
\label{swobservablesflow}
\end{eqnarray}
where $\Theta^n_p$ are given by (\ref{formsw}).

Follow the procedure we did in Eq. (\ref{veinticinco}), it is an
easy matter to check that these asymptotic observables
$\widetilde{\mathcal{O}}_{{\bf Y}_p}(\mu_p)$ are ${\cal
Q}$-invariant (BRST).

For an oriented manifold $M$ with $p_j$-vectors fields ${\bf
Y}_{p_j}$ ($p_j=0,1,2$), with $\sum_{j = 1}^r p_j = \rm{d}
(\mathcal{M}_{SW})$ and probability invariant measure, the $r$-point
correlation functions for the flow generated ${\bf Y}_{p_j}$ and
$\mu_{p_j}$ are given by

\begin{eqnarray}
\bigg\langle\widetilde{\mathcal{O}}^n_{{\bf Y}_{p_1}} (\mu_{p_1})
\cdots \widetilde{\mathcal{O}}^n_{{\bf Y}_{p_r}}
(\mu_{p_r})\bigg\rangle = \int (\mathcal{D} \mathcal{X}) \exp (-
{L}_{SW}/e^2 ) \prod_{j=1}^r \int_M i_{{\bf Y}_{p_j}}
(\Theta^n_{p_j}) \mu_{p_j} . \nonumber \\
\end{eqnarray}
This expression is reduced to the ordinary Seiberg-Witten invariants
$(\ref{eq:inevitabletwo})$, when the measure is supported on the
cycles. This means if $\mu_p = \sum_{j=1}^r \mu_{p_j}$ where each
$\mu_p$ is distributed uniformly over the cycles of $M$.

Similarly to the Donaldson case let us assume that the only zero
modes correspond to the abelian gauge field $A_\mu$ and its
BRST-like companion $\psi_\mu$. Following a similar procedure as in
the Donaldson case to compute the partition function we get
\begin{eqnarray}
\bigg\langle \widetilde{\mathcal{O}}_{{\bf Y}_{p_j}} (\mu_{p_j})
\bigg\rangle & = & \int_\mathcal{M_{SW}} da_1 \ldots da_n d\psi^1
\ldots d\psi^n \widetilde{\Phi}_{i_1 \cdots  i_n}(a_i,{\bf Y}_{p_j})
\psi^{i_1} \ldots \psi^{i_n} \nonumber \\& = & \int_\mathcal{M_{SW}}
\widetilde{\Phi}_{{\bf Y}_{d({\cal M})}},
\end{eqnarray}
where $\widetilde{\Phi}_{{\bf Y}_{p_j}} (\mu_{p_j}) =
\widetilde\Phi_{i_1 \cdots  i_n}(a_i,{\bf Y}_{p_j}) \psi^{i_1}
\ldots \psi^{i_n}$, $a$'s are the zero modes of the gauge field and
$\psi$'s are the zero modes of the fermionic field and
$\widetilde\Phi (a, {\bf Y}_{p_j})$ is a function that only depends
of the zero modes of the gauge field and contains the information of
the flow.

We integrate out $a_i$'s and obtain a $n$-form $\widetilde{\Phi}$
defined in the moduli space. Now suppose that
$\widetilde{\mathcal{O}} = \widetilde{\mathcal{O}}_{{\bf Y}_{p_1}}
(\mu_{p_1}) \cdots \widetilde{\mathcal{O}}_{{\bf Y}_{p_r}}
(\mu_{p_r})$ with $\sum_{p,j} p_j = n= d({\cal M}_{SW})$ and $p_j$
is the number of zero modes of $\widetilde{\mathcal{O}}_{{\bf
Y}_{p_j}} (\mu_{p_j})$. These functions define forms in the moduli
space in the following way
\begin{eqnarray}\nonumber
 \bigg\langle\widetilde{\mathcal{O}}_{{\bf Y}_{p_1}}
(\mu_{p_1}) \cdots \widetilde{\mathcal{O}}_{{\bf Y}_{p_r}}
(\mu_{p_r})\bigg\rangle & = & \int_{{\mathcal{M}}_{SW}}
\widetilde{\Phi}_{{\bf Y}_{p_1}}
\wedge \cdots \wedge \widetilde{\Phi}_{{\bf Y}_{p_r}}\\
& = &  \int_{{\cal M}_{SW}} \widetilde{\nu}_{{\bf Y}_{p_1}} \wedge \cdots \wedge \widetilde{\nu}_{{\bf Y}_{p_r}}.
\end{eqnarray}

In the simply connected case $(\pi_1 (M) = 0)$, the important
observables are those associated with cycles of zero dimension
$\gamma_0$ and of dimension two $\gamma_2$. In general a
$k_\gamma$-cycle has associated an operator (form) with ghost number
$U= 2 - k_\gamma$, this is the analog of the Donaldson
map $H_k (M) \to H^{2-k} (\mathcal{M}_{SW})$. Finally it is easy to
see that for $k=2$ the product of $r$ operators yields
\begin{eqnarray} \nonumber
 \bigg\langle\widetilde{\mathcal{O}}_{{\bf Y}_{1}}
(\mu_1)\cdots \widetilde{\mathcal{O}}_{{\bf Y}_{d/2}}
(\mu_{d/2})\bigg\rangle & =  & \int_{{\cal M}_{SW}} \widetilde{\nu}_{{\bf Y}_{1}} \wedge \cdots \wedge \widetilde{\nu}_{{\bf Y}_{d/2}}\\
& = & \# \left(H_{ \widetilde{\Sigma}_{{\bf Y}_{1}}} \cap \ldots
\cap H_{\widetilde{\Sigma}_{{\bf Y}_{d/2}}} \right),
\label{swinvariants}
\end{eqnarray}
where we have assumed the notation ${\bf Y}_{2_j} = {\bf Y}_{j}$ and
$\mu_{2_j}=\mu_j$  for asymptotic 2-cycles.

These ideas can be generalized by considering a twisted version of
the Yang-Mills theory with non-abelian gauge group and this leads to
non-abelian monopole equations
\cite{Labastida:1995zj,Labastida:1995gp}. We consider the general
case, when $M$ is not Spin but a Spin$_c$ manifold. For  one
hypermultiplet the equations for a non-abelian connection $A_\mu$
coupled to a spinor $M_{\alpha} \in \Gamma(S^+ \otimes L^{1/2}
\otimes E)$, where $S^+$ is the spin bundle, $L^{1/2}$ is the
determinant line bundle of the Spin$_c$ structure and  $E$ is the
vector bundle associated to a principal $G$-bundle via some
representation of the gauge group. Then the equations for the moduli
space are given by
\begin{equation}
F^{+ a}_{\dot\alpha \dot\beta} + 4 i \overline{M}_{(\dot\alpha} (T^a) M_{\dot\beta)} = 0, \ \ \ \ \
(\nabla^{\alpha \dot\alpha}_E \overline{M}_{\dot\alpha}) = 0,
\end{equation}
$T^a$ are the generators of the Lie algebra, $\nabla^{\alpha
\dot\alpha}_E$ is the Dirac operator constructed with the covariant
derivative with respect to the gauge connection $A_\mu$. Thus it is
possible to extend the set of observables (\ref{swobservablesflow})
for the non-abelian case. Similar computations can be done for
obtaining the Seiberg-Witten invariants associated to non-abelian
monopoles \cite{Labastida:1995zj} for the case of compact K\"ahler
manifolds following Ref. \cite{Labastida:1995gp} and making a
similar procedure as described in Sec. 5.2.

\subsection{Relation to Donaldson Invariants}

The computation of Donaldson-Witten and Seiberg-Witten invariants
for K\"ahler manifolds can be obtained by physical methods. The
difference lies in the underlying physics of the dynamics of ${\cal
N}=2$ supersymmetric gauge theories. For the Donaldson-Witten case
it was needed to add a mass term to soft breaking supersymmetry.
This introduces a non-trivial canonical divisor defined as the zero
locus of the mass term. However for the Seiberg-Witten case this is
not necessary. The underlying dynamics describing the strong
coupling limit of the gauge theory was elucidated in \cite{SW}
through the implementation of S-duality. The dual theory was used
later by Witten in Ref. \cite{MONO} to find new invariants of four
manifolds, the Seiberg-Witten invariants. In that paper it was found
a relation between both invariants. This subsection contains the
description of this relation when there are non-singular global
flows on the manifold.

The relevant ingredients are the operators $\widetilde{I}_{{\bf
Y}_a}$ and ${\cal O}$ inserted in $M$. Then the Donaldson invariants
take the following form
$$
\bigg{\langle} {\exp} \bigg( \sum_a \alpha_a \widetilde{I}_{{\bf
Y}_a}(\mu_a) + \lambda {\cal O}\bigg) \bigg \rangle
$$
\begin{equation}
= 2^{1+{1\over 4}(7\chi + 11 \sigma)} \bigg[{\exp} \bigg( {1\over
2} \widetilde{v}^2 + 2 \lambda  \bigg)  \sum_{\widetilde{x}} \widetilde{SW}({\widetilde{x}}) e^{\widetilde{v} \cdot \widetilde{x}} +
i^{\Delta}{\exp} \bigg( -{1\over 2} \widetilde{v}^2 - 2 \lambda  \bigg) \cdot
\sum_{\widetilde{x}}\widetilde{SW}({\widetilde{x}})e^{-i\widetilde{v} \cdot \widetilde{x}}\bigg]
\end{equation}
where $\alpha_a$ and $\lambda$ are complex numbers,
\begin{equation}
\widetilde{v}^2 = \sum_{a,b} \alpha_a \alpha_b
\#(\widetilde{\Sigma}_{{\bf Y}_a} \cap \widetilde{\Sigma}_{{\bf
Y}_b}),
\end{equation}
\begin{equation}
\widetilde{v} \cdot \widetilde{x} = \sum_{a} \alpha_a
\#(\widetilde{\Sigma}_{{\bf Y}_a} \cap \widetilde x)
\end{equation}
and  $\widetilde{SW}$ is the asymptotic version of the
Seiberg-Witten invariant (\ref{swinvariants}).

Here $\# (\widetilde{\Sigma}_{{\bf Y}_a} \cap \widetilde{x})$ is the
asymptotic intersection number between $\widetilde{\Sigma}_{{\bf
Y}_a}$ and a $\widetilde{x}$ and it is given by
\begin{equation}
\# (\widetilde{\Sigma}_{{\bf Y}_a}\cap \widetilde{x} ) = \int_{M_1 \times M_2} (\Theta_2\wedge \eta_{{\bf Y}_a})(x_1) \wedge (x\wedge \eta_{\bf X})(x_2) \cdot \delta(x_1-x_2),
\label{intCwSthree}
\end{equation}
where $\eta_{{\bf Y}_a}$ and $\eta_{\bf X}$ are the Poicar\'e dual
of $\widetilde{\Sigma}_a$ and $\widetilde{x}$.

In the previous equations we have used the following definitions
$x=- 2 c_1(L\otimes L) \in H^2(M,\mathbb{Z})$ and
$\widetilde{x}_{{\bf X}} = \int_M i_{{\bf X}} (x)  \mu$ with ${\bf
X}$ being the 2-vector field wrapping $\widetilde{x}$.

\section{A Physical Interpretation}

In this paper we have assumed the existence of ``diffused" cycles in
a suitable four-manifold. Due their relevance in dynamical systems
on simply-connected 4-manifolds we have considered asymptotic
2-cycles $\widetilde{\Sigma}_{{\bf Y}_1}, \dots
,\widetilde{\Sigma}_{{\bf Y}_r}$ together with a set of invariant
probability measures $\mu = \mu_1 + \cdots + \mu_r$, with $\mu_i$
supported in $\widetilde{\Sigma}_{{\bf Y}_i}$ on $M$. The present
section is devoted to interpret this system in terms of string
theory. Thus one would wonder if these 2-cycles can be interpreted
as closed string probes propagating on the underlying four-manifold
(which would be compact or non-compact). For this issue there is a
nice response through the computation of the scattering amplitudes
of an axion at zero momentum with a NS5-brane in the heterotic
string theory \cite{Harvey:1991hq}. In the present paper we follow
this direction and we argue that Eq. (\ref{donaldsoninvs}) is a
consequence of these considerations.

Let us consider a spacetime manifold $M^{1,9}$ provided with a set
of Borel probability measures invariant under a non-singular smooth
2-flow generated by a 2-vector field $Y$. Moreover we take the
following splitting $M^{1,9}= M^{1,5} \times M$, where $M^{1,5}$ is
a flat Minkowski space and $M$ is the transverse space. We consider
a NS fivebrane (NS5) as a solitonic object filling the space
$M^{1,5}$. Thus the transverse space $M$ consists of a four-manifold
parameterizing the positions of the NS5-brane. Our flow 2-orbit can
be regarded as a closed string propagating on $M^{1,9}$ without
necessarily being  localized in a homology cycle of  $M^{1,9}$. For
the purposes of this paper we focus on a special situation by
limiting the 2-flow to be defined only in the transverse space $M$ and
supported in the whole $M$. Consequently the Borel measures will be
defined only on $M$. In this case we can think of the 2-flow as an
asymptotic cycle $\widetilde{\Sigma}_{{\bf Y}}$) representing a
``diffuse" closed string propagating in $M$ and viewing the NS5 as a
scattering center.  If the probability measures are totally on $M$
then the NS5-brane sector will remain unchanged and the moduli space
of instantons remain the same. The effective action described in
\cite{Harvey:1991hq} is a non-linear sigma model on the worldvolume
of the NS5 brane $W$ and with target space the space ${\cal M}_N$.
This latter space represents the space of static NS5 brane solutions
and it is equal to the moduli space of $N$ Yang-Mills instantons
${\cal M}_N(M)$ over $M$. Thus the ground states correspond to
cohomology classes on ${\cal M}_N(M)$.  If we identify the ``diffuse''
heterotic closed string with the asymptotic 2-cycles on $M$ i.e.
$\widetilde{\Sigma}_{{\bf Y}}$, then the action is given by:
\begin{equation}
S= \int_M i_{{\bf Y}} (B)  \mu.
\end{equation}
Then from the interacting terms coupling the $B$-field with the
gauginos of the heterotic supergravity action we have
\begin{equation}
\widetilde{O}_{ij,\bf Y}=\int_M i_{{\bf Y}} (Z_{ij})  \mu,
\end{equation}
where $Z={\rm tr} (\delta_i A \wedge \delta_j A - \phi_{ij} F)$.

Now we would like to consider multiple axion scattering with zero
momentum. Then the transitions among the quantum ground states of
the worldvolume $W$, for instance, from $|0\rangle$ to  $|m\rangle$,
induced by the scattering of $r$ axions with the NS5-branes is
described by the scattering amplitude. The $r$ axions represent $r$
closed heterotic string wrapping the homology cycles
$\widetilde{\Sigma}_{{\bf Y}_1}, \dots ,\widetilde{\Sigma}_{{\bf
Y}_r}$, associated with the 2-vector fields ${\bf Y}_1, \dots ,{\bf
Y}_r$. Thus the scattering amplitude is given by
$$
{\cal A}({\bf Y}_1, \cdots ,{\bf Y}_r) = \langle m| \widetilde{O}_{{\bf Y}_1}\cdots \widetilde{O}_{{\bf Y}_r}   | 0  \rangle
$$
\begin{equation}
= \int_{{\cal M}_N} \widetilde{O}_{{\bf Y}_1}\wedge \cdots \wedge \widetilde{O}_{{\bf Y}_r}.
\end{equation}
Thus we have deduced eq. (\ref{donaldsoninvs}) from string theory.

Above, we assumed that the set of Borel measures are distributed along the space $M$.
This is consistent with the fact we have in the transverse directions (along $M^{1,5}$)
filled with the NS5-brane which is a solitonic object and consequently is very heavy in
the perturbative regime and consequently they are very difficult to excite.

S-duality between the heterotic and type I string interchanges NS5
by a D5-brane and axions by D1-branes \cite{Polchinski:1995df}. The
self-dual gauge field on $M$ leads to the ADHM construction of
instantons \cite{Witten:1995gx}. The interactions are now given by
the Type I string action.  It would be interesting to make a
description of the asymptotic cycles within this context.

\vskip 1truecm
\section{Final Remarks}

In the present paper we look for the implementation of the procedure
followed in Ref. \cite{VV} for Jones-Witten invariants, to compute
invariants for flows in higher dimensional manifolds. In the this
situation the relevant invariants of interest were the smooth
invariants of four-manifolds i.e. the Donaldson-Witten and the
Seiberg-Witten invariants. We were able to obtain these invariants
when some flows generated by non-singular and non-divergence-free
smooth $p$-vector fields are globally defined on the four-manifold.
We assumed that the homology cycles of $M$ are described by the
asymptotic cycles. We focus our work on simply-connected
four-manifolds, thus the only relevant flows are the 2-flows, though
the invariants can be defined for any other $p$-flows. This is the
situation that leads to a generalization of invariants of
four-manifolds with flows.

In order to implement the above considerations we use the Witten
cohomological field theory, whose observables are cohomology classes
of $M$. In the presence of flows these observables were constructed
as geometric currents underlying asymptotic cycles and foliations
introduced by Ruelle and Sullivan \cite{RS} and Schwartzman
\cite{SH}. Thus the asymptotic observables and their correlation
functions give rise to new smooth invariants for four-manifolds with
a dynamical system with an invariant probability measure. That is,
they represent smooth invariants for foliations i.e. triples
$(M,{\cal F},\mu)$. This was done for Donaldson-Witten invariants
(\ref{donaldsoninvs}) as well as for Seiberg-Witten invariants
(\ref{swinvariants}). Donaldson-Witten invariants are also obtained
for the case of K\"ahler manifolds with flows and some examples were
described in Sec. 5.3. Finally, we attempt to give a physical
interpretation in terms of string theory. We used the procedure
outlined in Ref. \cite{Harvey:1991hq} to obtain the invariants
(\ref{donaldsoninvs}) as scattering amplitudes of $r$ axions at the
zero momentum with $N$ coincident NS5-branes in the heterotic string
theory. These axions are of special character and they represent $r$
2-flows wrapping $r$ homology 2-cycles of $M$. We gave only general
remarks on this subject and a further detailed analysis must be
performed. This includes the incorporation of S-duality between the
heterotic and Type I string theory \cite{Polchinski:1995df} and the
uses of this structure \cite{Witten:1995gx} to construct an ADHM
construction of instantons with flows. It would be interesting to
include flows in terms of proper dynamical fields such as is the
case in string theory such as the NS $B$-field and the RR fields and
make similar consideration as the present paper, but this time, in
terms of specific interactions of the flow degrees of freedom. In
this case it is possible to compute the back reaction of the fields
of the theory to the flow.

There is a number of possible further generalizations of our work.
One of them is the extension of asymptotic invariants to quantum
cohomology \cite{TSM,MM,Kontsevich:1994qz}, by considering
asymptotic cycles in the target space of a topological non-linear
sigma model of types $A$ or $B$ \cite{MM}. Of special interest is
the possibility to define an asymptotic version of the
Rozansky-Witten invariants \cite{Rozansky:1996bq}. This is due to
the fact that their construction involves a topological sigma model
on a 3-manifold and target space being an hyper-K\"ahler manifold.
The theory leads to link-invariants on the 3-manifold with
underlying structure group labeled by the hyper-K\"ahler structure.
We would like to establish a relation with the results obtained in
Ref. \cite{VV}.

Another possibility is the consideration of flows generated by
$p$-vector fields on supermanifolds. The analysis involves the
computation of correlation functions with even and odd operators.
This will constitute a supersymmetric extension of the work
considered in the present paper. One more possible direction
constitutes the implementation of the procedure to the computation
of correlation functions of observables in the eight-dimensional
generalization of the cohomological field theory
\cite{Baulieu:1997jx,Acharya:1997jn}. Some of these issues are
already under current investigation and will be reported elsewhere.

\vskip 2truecm
\centerline{\bf Acknowledgments}

We would like to thank the referee for carefully reading our
manuscript and for giving us very important suggestions. The work of
H. G.-C. is supported in part by the CONACyT grant 128761. The work
of R.S. was supported by a CONACyT graduate fellowship. The work of
A.V. was partially supported by CONACyT grant number 129280 and
DGAPA-PAPIIT, Universidad Nacional Aut\'onoma de M\'exico.



\vskip 1truecm

\addcontentsline{toc}{section}{Bibliography}
\bibliographystyle{alpha}



\end{document}